\documentclass[pra,amsmath,amssymb,aps,showpacs, twocolumn,superscriptaddress]{revtex4-1}

\usepackage{amsfonts}
\usepackage[T1]{fontenc}

\usepackage{graphicx}
\usepackage{amsmath}
\usepackage{float} 
\usepackage{amssymb}
\usepackage{color}
\usepackage[version=3]{mhchem}
\usepackage[colorlinks,bookmarks=false,citecolor=blue,linkcolor=red,urlcolor=blue]{hyperref}
\usepackage{hyperref}
\usepackage{url}
\newcommand{\ket}[1]{\mbox{$ | #1 \rangle $}}
\newcommand{\bra}[1]{\mbox{$ \langle #1 | $}}

\usepackage{multirow}

\usepackage{placeins}

\hypersetup{
colorlinks=true,
linkcolor=blue,         
citecolor=blue,         
filecolor=magenta,      
urlcolor=red
}

\newcommand\figwidthdouble{1.75} 
\newcommand\figwidthsingle{0.93} 

\begin{document}
\title{Spectral Structure and Many-Body 
Dynamics\\  of Ultracold Bosons in a Double-Well}

\author{Frank Sch\"afer}
\email[ ]{frank.schaefer@unibas.ch}
\affiliation{Physikalisches Institut, Albert-Ludwigs-Universit\"{a}t  Freiburg, Hermann-Herder-Stra{\ss}e 3, D-79104, Freiburg, Federal Republic of Germany}
\affiliation{Department of Physics, University of Basel, Klingelbergstrasse 82, CH-4056 Basel, Switzerland}
\author{Miguel A. Bastarrachea-Magnani}
\affiliation{Physikalisches Institut, Albert-Ludwigs-Universit\"{a}t  Freiburg, Hermann-Herder-Stra{\ss}e 3, D-79104, Freiburg, Federal Republic of Germany}
\affiliation{Department of Physics and Astronomy, Aarhus University, Ny Munkegade, DK-8000 Aarhus C, Denmark.}
\author{Axel U. J. Lode}
\affiliation{Physikalisches Institut, Albert-Ludwigs-Universit\"{a}t  Freiburg, Hermann-Herder-Stra{\ss}e 3, D-79104, Freiburg, Federal Republic of Germany}
\author{Laurent de Forges de Parny}
\affiliation{Physikalisches Institut, Albert-Ludwigs-Universit\"{a}t  Freiburg, Hermann-Herder-Stra{\ss}e 3, D-79104, Freiburg, Federal Republic of Germany}
\affiliation{ACRI-ST, 260 route du Pin Montard, 06904 Sophia Antipolis CEDEX, France}
\author{Andreas Buchleitner}
\email[ ]{andreas.buchleitner@physik.uni-freiburg.de}
\affiliation{Physikalisches Institut, Albert-Ludwigs-Universit\"{a}t  Freiburg, Hermann-Herder-Stra{\ss}e 3, D-79104, Freiburg, Federal Republic of Germany}
\affiliation{Freiburg Institute for Advanced Studies (FRIAS), Albert-Ludwigs-Universit\"{a}t  Freiburg, Albertstr. 19, D-79104 Freiburg, 
Federal Republic of Germany}

\date{\today}

\begin{abstract}
We  examine the spectral structure and many-body dynamics of two and three repulsively interacting bosons  trapped in a one-dimensional double-well, 
for variable barrier height, inter-particle interaction strength, and initial conditions. By exact diagonalization of the many-particle Hamiltonian, we specifically explore the dynamical behaviour of the particles launched either at the single particle ground state or saddle point energy, in a time-independent potential. We complement these results by a characterisation of the cross-over from diabatic to quasi-adiabatic evolution under finite-time switching of the potential barrier, via the associated time-evolution of a single particle's von Neumann entropy. This is achieved with the help of the multiconfigurational time-dependent Hartree method for indistinguishable particles (\textsc{Mctdh-x}) -- which also allows us to extrapolate our results for increasing particle numbers.
\end{abstract}


\maketitle

\section{Introduction}
\label{sec_1}

The detailed microscopic understanding of interacting many-particle quantum dynamics in state-of-the-art experiments with ultracold atoms \cite{serwane2011,jochim2015,Morsch_Oberthaler_RevMod_2006, Albiez_Oberthaler_PRL_2005, esslinger2008, Greiner_Bloch_2002, Bloch_2005, Bloch_Zwerger_2008, Bloch_Nature_2008, Gross_Bloch_2017} in well-characterised potential landscapes remains a challenging task for theory: While a large arsenal of advanced numerical techniques has been developed over the past two decades to efficiently simulate interacting many-particle dynamics \cite{schollwock2005, schollwock2011, wall2012, cederbaum_MCTDHX, lode2019RMP}, all of them 
must ultimately surrender when confronted with truly complex dynamics, i.e., under conditions where a generic initial state fully explores, on sufficiently 
long time scales, an exponentially large Hilbert space in the number of particles and/or degrees of freedom. By the very meaning of complexity, even the most efficient numerical 
methods can only be expected to yield reliable results when the dynamics can be restricted to finite sub-spaces of the exponentially large Hilbert spaces 
-- either by reducing the time window over which the evolution is followed, or by choosing physical situations which a priori confine the many-particle state.
This has been long understood in the light-matter interaction of atoms and molecules \cite{parker2003}, as well as in quantum chaos \cite{abu2003}, and meets revived interest 
given the experimental progresses in the control of cold matter \cite{pasek2017}. 

While it is therefore clear that the only promising route for an efficient characterisation of large and complex quantum systems can be through effective descriptions
-- such as offered, e.g., by the theory of open quantum systems \cite{davies1976, alicki2007, gardiner2004, breuer2002}, modern semiclassics \cite{schlagheck2019}, 
or random matrix theory \cite{guhr1998, walschaers2016} -- there is an intermediate range of system sizes 
where efficient numerical methods can a) be gauged against each other, to benchmark their quantitative reliability, without any a priori restriction on the explored 
portion of Hilbert space, and b) contribute to gauge effective theories against (numerically) exact 
solutions \cite{lindinger2019, MCTDHX_Lode_Streltsov_2012, fasshauer2016multiconfigurational}, at spectral densities where quantum granular effects induce 
possibly sizeable deviations \cite{schlawin2016} from effective theory predictions (which always rely on some level of coarse graining).
In our view, it is this intermediate system sizes where efficient methods of numerical simulation develop their full potential, since they can inspire and ease the development, e.g.,
of powerful statistical methods and paradigms (such as scaling properties \cite{lindinger2019, carnio2019, pasek2017}) -- which then enable robust predictions in the realm of fully unfolding complexity.

In the present paper, we contribute to this line of research by exploring the spectral and dynamical properties of few bosonic particles loaded into a symmetric double-well potential, with static or switchable tunneling barrier. 
Prima facie, this is a well-known and text-book-like example, yet with a panoply of experimental realisations, and of paradigmatic relevance as an incarnation of Josephson dynamics \cite{Albiez_Oberthaler_PRL_2005, Folling_Bloch_Nature_2007,Milburn_Walls_1997, Smerzi_Shenoy_1997, Menotti_2001, Mahmud_Reinhardt_2005, Salgueiro_Weidemuller_2007, Murphy_Busch_2007, Murphy_McCann_2008, cederbaum_splitting_dynamics, Sakmann_Cederbaum_PRL_2009, zollner_Schmelcher_PRA_2006, Zollner_Meyer_Schmelcher_PRA_2007, zollner_Schmelcher_PRA_2008, zollner_few_bosons, Dobrzyniecki_Sowinski_EPJD2016, Spagnolli} or as the elementary building block of quantum dynamics in lattice-like structures \cite{parra2013}, and quickly defines a formidable numerical challenge if only one admits excitations far beyond the immediate vicinity of the ground state energy, and seeks to accurately monitor the long-time dynamics of two or more particles.
We will see how the spectral structure of the single-particle problem is amended by adding a second, identical particle, and how finite-strength interactions restructure the many-particle spectrum and eigenstates, throughout the excitation spectrum up to the vicinity of the potential barrier. 

Here and in the following, we use the term ``many-body/particle'', albeit the systems we consider are composed of a relatively small number of particles.
Note that our considerations are from first principles and start from the many-body Hamiltonian. 
Moreover, it has been shown theoretically \cite{abu2003, kolovsky2003, kolovsky2004} and experimentally \cite{serwane2011,jochim2015} that the physics of interacting few-body systems can very quickly approach the many-body limit.

The spectral information thus generated allows us to decipher characteristic features of the many-particle dynamics, for distinct choices of the initial condition, and over a wide range of interaction strengths, for static as well as for diabatically or (quasi-)adiabatically ramped 
potential barriers. Finally, we illustrate, through an analysis of the von Neumann entropy of the (reduced) single-particle density matrix, how such transition from diabatic to (quasi-) adiabatic switching controls the effectively explored sub-volume of Hilbert space, and how robust coarse grained features of the resulting ``phase diagram'' emerge  as the particle number is increased from two to ten. 
The latter case can only be treated with the help of the \textsc{Mctdh-x} \cite{lode2016, fasshauer2016multiconfigurational, softwareMCTDHX} method which has been verified against exact \cite{MCTDHX_Lode_Streltsov_2012, fasshauer2016multiconfigurational}
and experimental \cite{nguyen2019param} results and is reviewed in Ref.~\cite{lode2019RMP}. Here, we push \textsc{Mctdh-x} to its limits in monitoring long-time dynamics of rather moderate, mesoscopic particle numbers, in the presence of strong, switching-induced excitations (``quenches'').

The paper is organized as follows:  The theoretical framework, including a brief description of the numerical methods, 
is presented in Sec.~\ref{sec_2}.  Section~\ref{sec_3} is devoted to the discussion of 
the spectral and eigenstate structure of the problem at hand.
First, Sec.~\ref{sec3_subA} discusses how the energy spectrum depends on both the tunneling barrier height 
and the inter-particle interaction strength, for two and three particles. 
Next, in Sec.~\ref{sec3_subB}, we study few-body correlations encoded in the few-body eigenstates. 
This prepares our analysis 
of the dynamics in Section \ref{sec4}.
In Sec.  \ref{sec4_subA}, we investigate the dynamics of two particles in a static double-well potential, initially prepared in two different states: A superposition of low-lying states, and a superposition of excited states with energies close to the saddle-point.
Finally, we consider the  scenario of a time-dependent potential in Sec.~\ref{sec4_subB}: With the atoms initially prepared in the ground state of a harmonic trap, a central 
barrier is ramped-up, and the thereby induced dynamics can be tuned from diabatic to (quasi-) adiabatic by appropriate control of the ramping time. 
Our results are summarized in Sec. \ref{sec_5}.

\section{Hamiltonian and Methods}
\label{sec_2}

\subsection{Hamiltonian of trapped interacting bosons}
\label{sec2_subA}

The Hamiltonian of $N$ spinless, ultracold atoms with repulsive contact interaction and confined to a one-dimensional double-well potential reads in atomic units

\begin{eqnarray}
  H &=& \sum_{i}^N  \left (- \frac{1}{2}  \frac{ d^2 }{d x^2_i} +V(x_i,t) \right ) + \frac{\lambda}{2} \sum_{i\neq j} \delta(x_i-x_j)~, \ \ 
\label{MB_Hamiltonian}
\end{eqnarray}

where
\begin{eqnarray}
V(x_i,t)=\frac{x_i^2}{2}+A(t)e^{-x_i^2/2} \ \ 
\label{Double_Well_Potential}
\end{eqnarray}
allows for a non-trivial time-dependence of the potential barrier, through the time-dependence of $A(t)$, 
$x_i$ denotes the position of the $i$th particle, and the repulsive interaction strength $\lambda>0$ is determined by  
the s-wave scattering length  and the transverse confinement \cite{Olshanii_1998}.

The minimum of $V(x_i,t)$ is located at $x=0$ if $A(t)<1$ (single-well), or at $ x=\pm \sqrt{2 \ln{(A(t))}}$ if $A(t) \geq 1$ (double-well).
Both, static and time-dependent barriers will be considered. 
In the static case, the central barrier amplitude is constant,  $A(t)=A_{\rm max}$, whereas
in the  time-dependent scenario, the amplitude  is ramped up linearly according to
\begin{equation}
A(t)= A_{\rm max} \times
\left\{
\begin{array}{lll}
t / T_{\rm ramp}, \hspace{0.4 cm}     t<T_{ \rm ramp},  \\
1,    \hspace{1.2 cm}  \ \  t\geq T_{\rm ramp} .\\
\end{array}
\right.
\vspace{0.1 cm}
\label{Double_Well_Amplitude}
\end{equation}

\subsection{Numerical methods and observables}
\label{sec2_subB}

The spectral and dynamical properties of the Hamiltonian \eqref{MB_Hamiltonian} are numerically investigated by using three approaches: the Fourier Grid Hamiltonian (FGH), the Bose-Hubbard (BH) representation of a continuous  potential, and  the multiconfigurational time-dependent 
Hartree method for indistinguishable particles (\textsc{Mctdh-x}); see Appendices~\ref{Method_FGH}, \ref{Method_BH}, and \ref{Method_MCTDH}, respectively.

Each of these is suited for a specific task.
We use FGH and BH which, ultimately, rely on different basis set representations of the Hamiltonian, to infer the spectrum   
of  $N\leq2$ and $N=3$ interacting bosons, 
by direct diagonalization.
FGH is also useful for the investigation of the quenched dynamics when a harmonic potential with $A_{\rm max}=0$ at $t=0$ is suddenly transformed  into a static 
double-well with fixed barrier $A_{\rm max}=const.$ at $t>0$ [in other words, $T_{ \rm ramp}\rightarrow 0$ in Eq.~\eqref{Double_Well_Amplitude}]. 
For our study of the case  of $N \leq 10$ interacting bosons in a time-dependent double-well with $T_{ \rm ramp}\neq0$, we use the \textsc{Mctdh-x} method which enables accurate results for the dynamics, but cannot provide the complete spectral information as the BH/FGH methods. Since dynamical properties of interacting many-particle systems emerge, already at rather small particle numbers \cite{abu2003}, the combination of all three approaches can be considered complementary.

FGH and BH yield the $N$-particle eigenenergies
\begin{eqnarray}
 H \ket{\Psi_n}=E_n^{NP}(A_{\max}, \lambda)\ket{\Psi_n}  ~,
\label{eigenenergies_Nparticles}
\end{eqnarray}
with $\ket{\Psi_n}$ the $N$-particle  eigenvector with quantum number $n$. All eigenstates are normalized to unity, throughout this paper.
The  quantity 
\begin{eqnarray}
\left|\psi_n(x_1,x_2, \dots, x_N) \right|^2= |\langle x_1, \dots, x_n  \ket{\Psi_n} |^2
\label{probability_densities}
\end{eqnarray}
yields the associated probability density to find $N$ bosons located at positions $x_1, x_2, \dots, x_N$, respectively.
Visualizations thereof reflect the correlations between the positions of the
particles \cite{zollner_few_bosons, zollner_Schmelcher_PRA_2008, Sakmann_Cederbaum_PRL_2009, zollner_Schmelcher_PRA_2006, Murphy_Busch_2007, Murphy_McCann_2008, Hunn_2013, FS_thesis_2018, nguyen2019param}, 
which can be assessed, e.g., through their entanglement. 
A possible (though certainly non-exhaustive) quantifier of the 
non-separability of 
a general many-particle state $\ket{\Psi(t)}$ is given by the von Neumann  entropy
\begin{equation}
S(t)=-\text{Tr}\left[\rho^{1P}(t)\ln \rho^{1P}(t)\right]
\label{Von_Neumann_Entropy}
\end{equation}
of the reduced single-particle density matrix 
\cite{Mack_Freyberger_2002, Sun_2006, Sowinski_Rzazewski_2010, Murphy_Busch_2007, Murphy_McCann_2008},
where $\rho^{1P}(x,x',t)$ is defined as the trace over all degrees of 
freedom of all but one boson of the full density operator, i.e.,
\begin{equation}
\rho^{1P}(t)=\text{Tr}_{2,\dots,N}\left[\ket{\Psi(t)}\bra{\Psi(t)}\right]~.
\label{RSPDM}
\end{equation}
In particular, $S=0$ if the state is \textit{separable}, while large values of $S$ are a hallmark of a strongly \textit{entangled} 
many-particle state \cite{Ghirardi_2003,Ghirardi_2004,ghirardi2004criteria,benatti2011,tichy2013}.

To characterize the dynamics of two bosons, we monitor the time-evolution of the particles' probabilities to reside both in the right (RR) or left (LL) well, or of each occupying one well (LR), given by \cite{Hunn_2013}
\begin{align}
P_{(LL)}(t) = &\int_{x_{\rm min}}^{0} \text{d}x_1\int_{x_{\rm min}}^{0} \text{d}x_2 ~ |\psi(x_1,x_2;t)|^2, \notag \\
P_{(RR)}(t) =& \int_{0}^{x_{\rm max}} \text{d}x_1 \int_{0}^{x_{\rm max}} \text{d}x_2~|\psi(x_1,x_2;t)|^2, \label{eq:detection-probabilities}\\
P_{(LR)}(t) =& ~2 \cdot \int_{x_{\rm min}}^{0} \text{d}x_1 \int_{0}^{x_{\rm max}} \text{d}x_2 ~|\psi(x_1,x_2;t)|^2,\notag 
\end{align} 
where we defined the three mutually distinct domains ${(LL)=(x_1<0,x_2<0)}$, ${(RR)=(x_1>0, x_2>0)}$, and ${(LR)=(x_1<0, x_2>0)\vee(x_1>0, x_2<0)}$. We also introduced
the minimum ($x_{\rm min}$) and maximum ($x_{\rm max}$) values of the grid in configuration space employed in the numerical approaches.
In addition, we evaluate the time-integrated probability current
\begin{widetext}
\begin{equation}
J_{(RR\to LR)}(t) = 2 \cdot  \int_{0}^{t} \text{d}t' \int_{0}^{x_{\rm max}} \text{d}x_2  ~\text{Im}\left[\psi^*(x_1,x_2;t')\frac{\partial}{\partial x_1}\psi(x_1,x_2;t')\right]_{x_1=0},
\label{eq:josephson-probabilityflux}
\end{equation}
\end{widetext}
where the factor 2 accounts for the bosonic symmetry. 
$J_{(RR\to LR)}$ is derived \cite{sh_diss} from the continuity equation and 
measures the probability flux within a time interval $t$ from domain $(RR)$ to domain $(LR)$.
This quantity is particularly important to distinguish first-order pairwise tunneling $(RR \to LL)$ from second-order pairwise tunneling $(RR\to LR \to LL)$. 
First-order, pairwise tunneling was observed \cite{sh_diss}, e.g., for attractively interacting bosons in a double-well, where $J_{(RR\to LR)}(t)=0, \forall t,$ when the particles are initially prepared in one well.   

\section{Structure of spectrum and eigenstates}
\label{sec_3}

\subsection{Few-body excitation spectra}
\label{sec3_subA}

Since the dynamics of the system is ultimately encoded in its spectrum, 
we first discuss  the  parametric evolution of the eigenvalues (\ref{eigenenergies_Nparticles})  of $N=1, 2$ and  $3$ bosons  with both the central barrier height $A_{\rm max}$ and the interaction strength $\lambda$.

\begin{figure*}[t]
\includegraphics[width=\figwidthsingle \columnwidth]{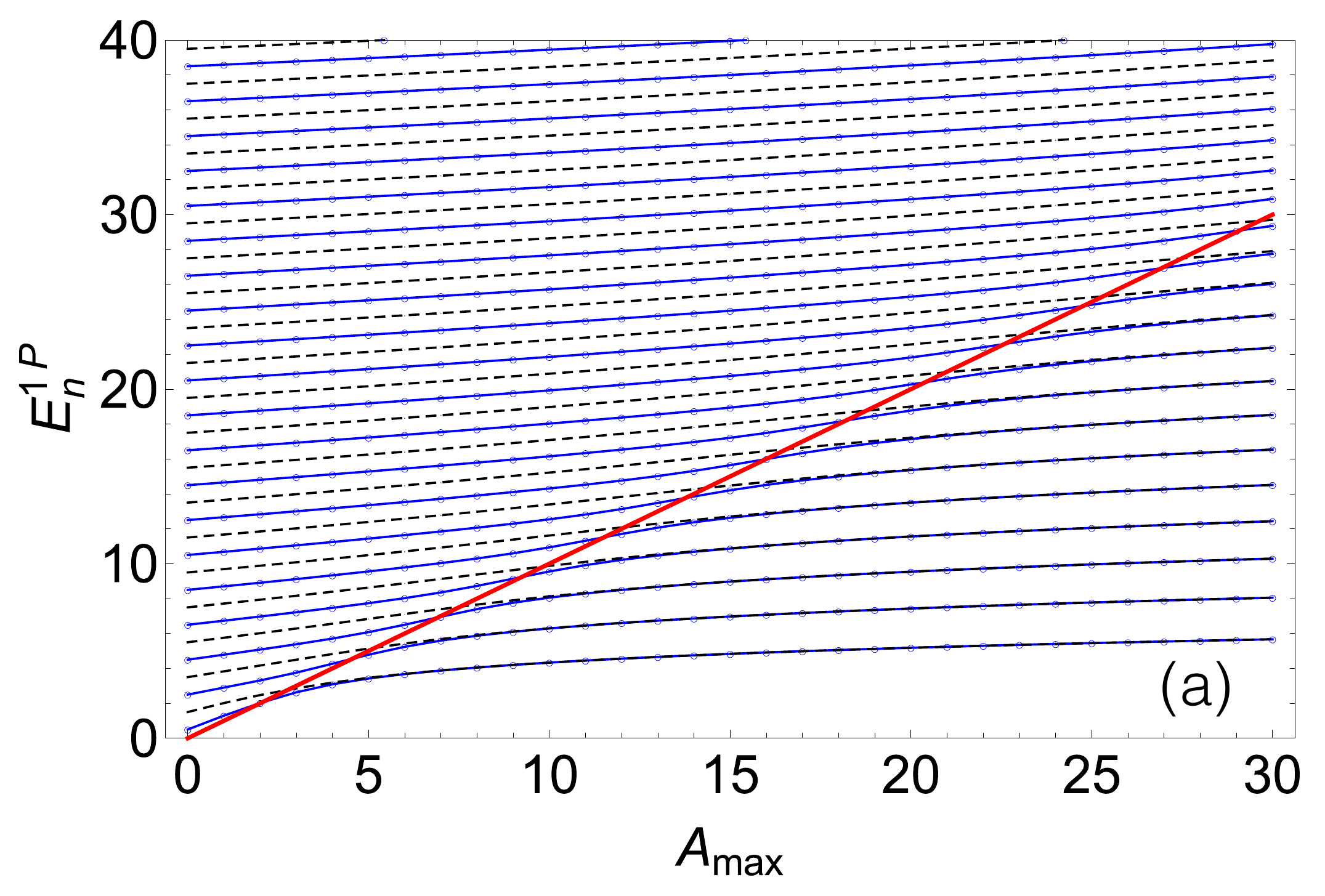}
\includegraphics[width=\figwidthsingle \columnwidth]{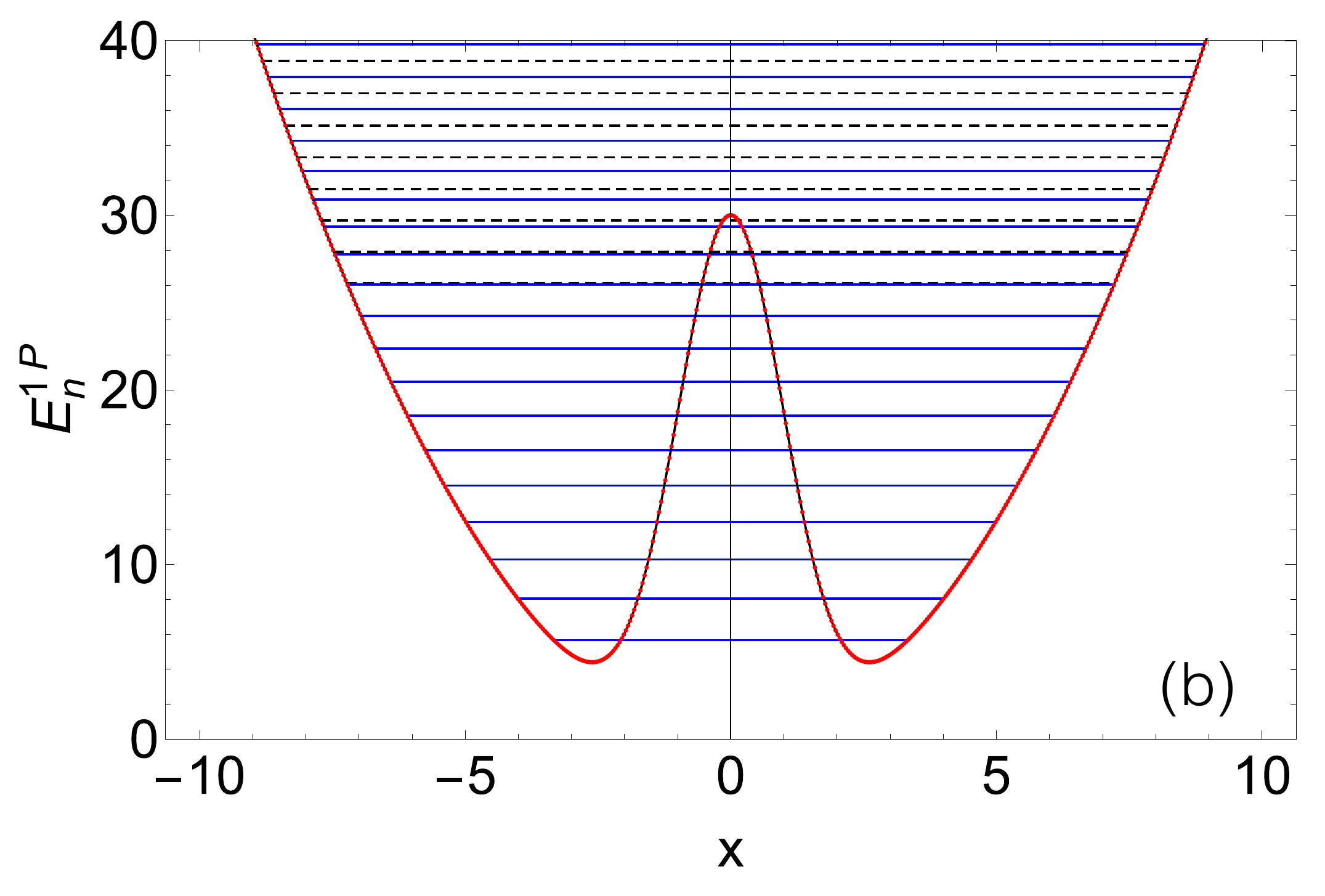}
\caption{(Color online) Single-particle eigenenergies $E_n^{1P}$ of Eq.~(\ref{1P_Hamiltonian}), (a) as a function of the tunneling barrier height $A_{\rm max}$, 
and (b) for $A_{\rm max}=30$ in the double-well potential (red). The red line in (a) indicates the central barrier's height on the energy axis.
Even- (blue lines) and odd-parity  (black dashed) states become nearly degenerate as $A_{\rm max}$
is increased.
Employed parameter values for the FGH method (see App.~\ref{Method_FGH}): $x_{\rm max}=-x_{\rm min}=40$, $N_{\rm cut}=330$, and $N_{\rm Grid}=2047$.
}
 \label{Figure1}   
\end{figure*}

The single-particle spectrum is obtained by solving the time-independent Schr\"odinger equation
\begin{eqnarray}
 \left (- \frac{1}{2}  \frac{ d^2 }{d x^2} +\frac{x^2}{2}+A_{\rm max}e^{-x^2/2}  \right ) \psi_n(x) = E_n^{1P} \psi_n(x). \ \ \  \ \ 
\label{1P_Hamiltonian}
\end{eqnarray}
Figure~\ref{Figure1} shows the evolution of the single-particle eigenenergies $E_n^{1P}$ as the central barrier height $A_{\max}$ is continuously increased from a harmonic trap ($A_{\max}=0$) to a deep double-well ($A_{\max}=30$).
In the harmonic limit,  the spectrum exhibits the well-known harmonic oscillator structure $E_n^{1P} (A_{\max}=0)=n+1/2$.
As the eigenenergies dive into the region below the barrier $A_{\max}$ (indicated by the red diagonal in Fig.~\ref{Figure1} (a)), the odd and even harmonic oscillator states become (nearly) degenerate.  
Sufficiently above $A_{\rm max}$, the energies are only weakly perturbed by the central barrier and we essentially recover the harmonic oscillator energy levels.  
In the limit $A_{\rm max}\to \infty$, the two wells decouple, leading to a fully degenerate harmonic oscillator spectrum.

From the structure of the single-particle spectrum, we can already anticipate that different dynamical behaviours can be expected for initial conditions with energies chosen below or above  $A_{\max}$, as will be elaborated upon, subsequently.

We now turn our attention to the spectrum of two particles obtained with the FGH method. 
The exact two-body spectrum is calculated by diagonalization of Eq.~\eqref{MB_Hamiltonian} represented in the single-particle basis,
as explained in Appendix~\ref{Method_FGH}. Figure~\ref{Figure2} (a) shows that, for  $A_{\max}=0$, we recover the well-known spectrum of two non-interacting bosons in a harmonic trap, i.e., 
$E_n^{2P}  (A_{\max}=0)=n+1$, with $n=n_1+n_2$ and degeneracy $g=n/2+1$ ($g=(n+1)/2$) for even (odd) $n \geq0$.
Here again, raising the central barrier gradually introduces a further degeneracy in the spectrum:
The first three lowest-lying states become (nearly)  degenerate when increasing $A_{\max}$.
This effect, also discussed in Ref.~\cite{Murphy_McCann_2008}, is a direct consequence of the twofold degeneracy of the single-particle ground-state of the double well, since all the eigenstates $| \Psi_0 \Psi_0 \rangle $,  $| \Psi_0 \Psi_1 \rangle $ and $| \Psi_1 \Psi_1 \rangle $, with  
\begin{equation}
| \Psi_n \Psi_m \rangle \equiv \frac{ | \Psi_n \rangle \otimes  | \Psi_m  \rangle   +   | \Psi_m \rangle \otimes  | \Psi_n  \rangle  }{\sqrt{2} \sqrt{1+ \langle \Psi_n | \Psi_m \rangle }} ~,
\label{2P_exact_diagonal_basis_main}
\end{equation}
acquire the same energy value at large $A_{\max}$ [see Eqs.~\eqref{2P_exact_diagonal_basis}~and~\eqref{FGH_one_body_diagonal_terms}].
For higher excitations, an analogous effect is observed: E.g., the energies of the states $| \Psi_2 \Psi_0 \rangle $, $| \Psi_2 \Psi_1 \rangle $, $| \Psi_3 \Psi_0 \rangle $ and $| \Psi_3 \Psi_1 \rangle $,  respectively given by the sums of single particle energies, $E_2^{1P}+E_0^{1P}$, $E_2^{1P}+E_1^{1P}$, $E_3^{1P}+E_0^{1P}$, and $E_3^{1P}+E_1^{1P}$,
converge  when increasing $A_{\max}$, since $E_1^{1P}\simeq E_0^{1P}$ and $E_3^{1P}\simeq E_2^{1P}$.
Therefore, the entire spectrum of two non-interacting bosons, plotted in
Fig.~\ref{Figure2} (a),
can be understood solely in terms of the single-particle spectrum.
The emergence of the sequence of quasi-degenerate states is clearly observed below the separatrix $E^{2P}=2A_{\max}$, plotted in red in 
Fig.~\ref{Figure2} (a).

\begin{figure*}[t]
\includegraphics[width=\figwidthsingle \columnwidth]{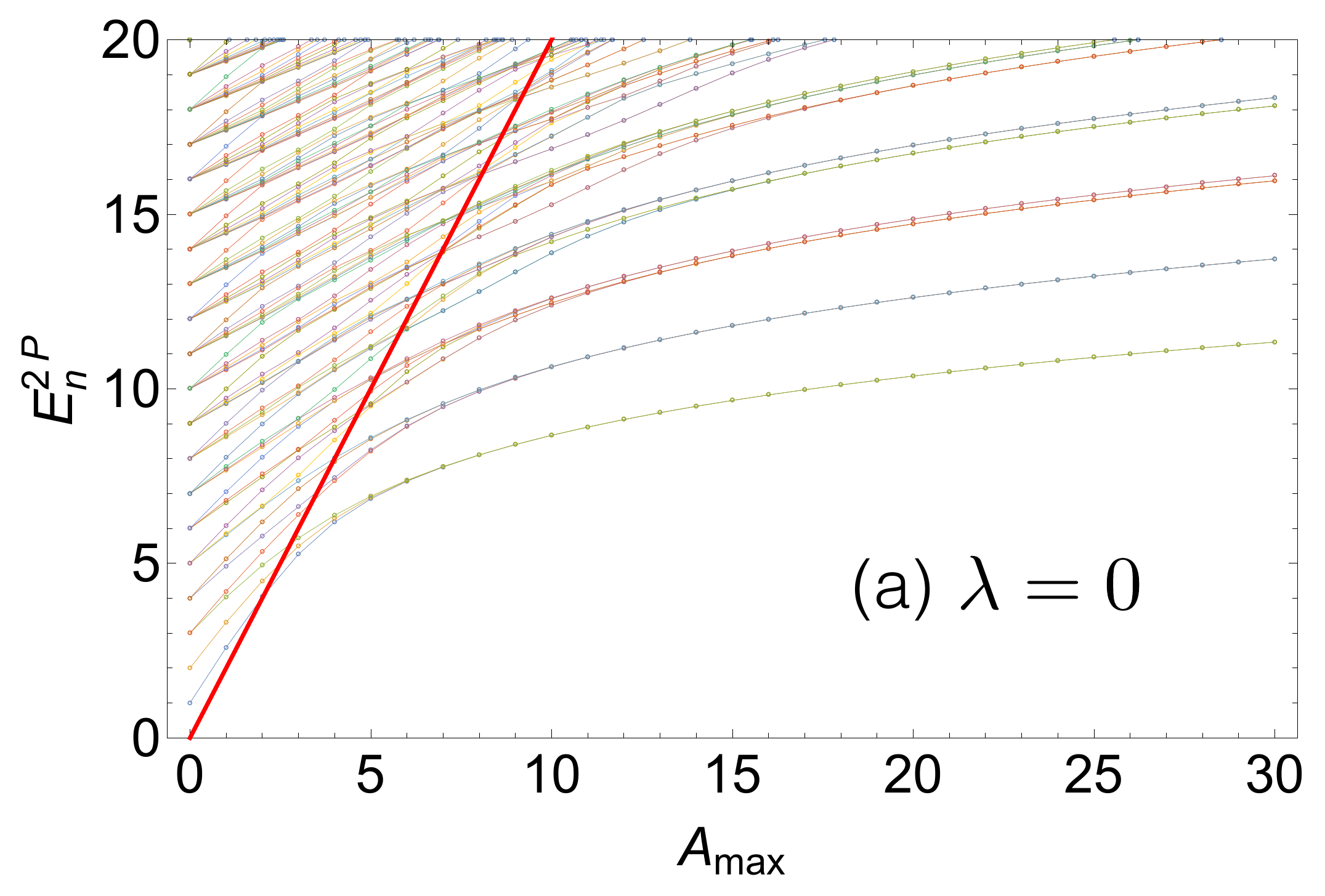}
\includegraphics[width=\figwidthsingle \columnwidth]{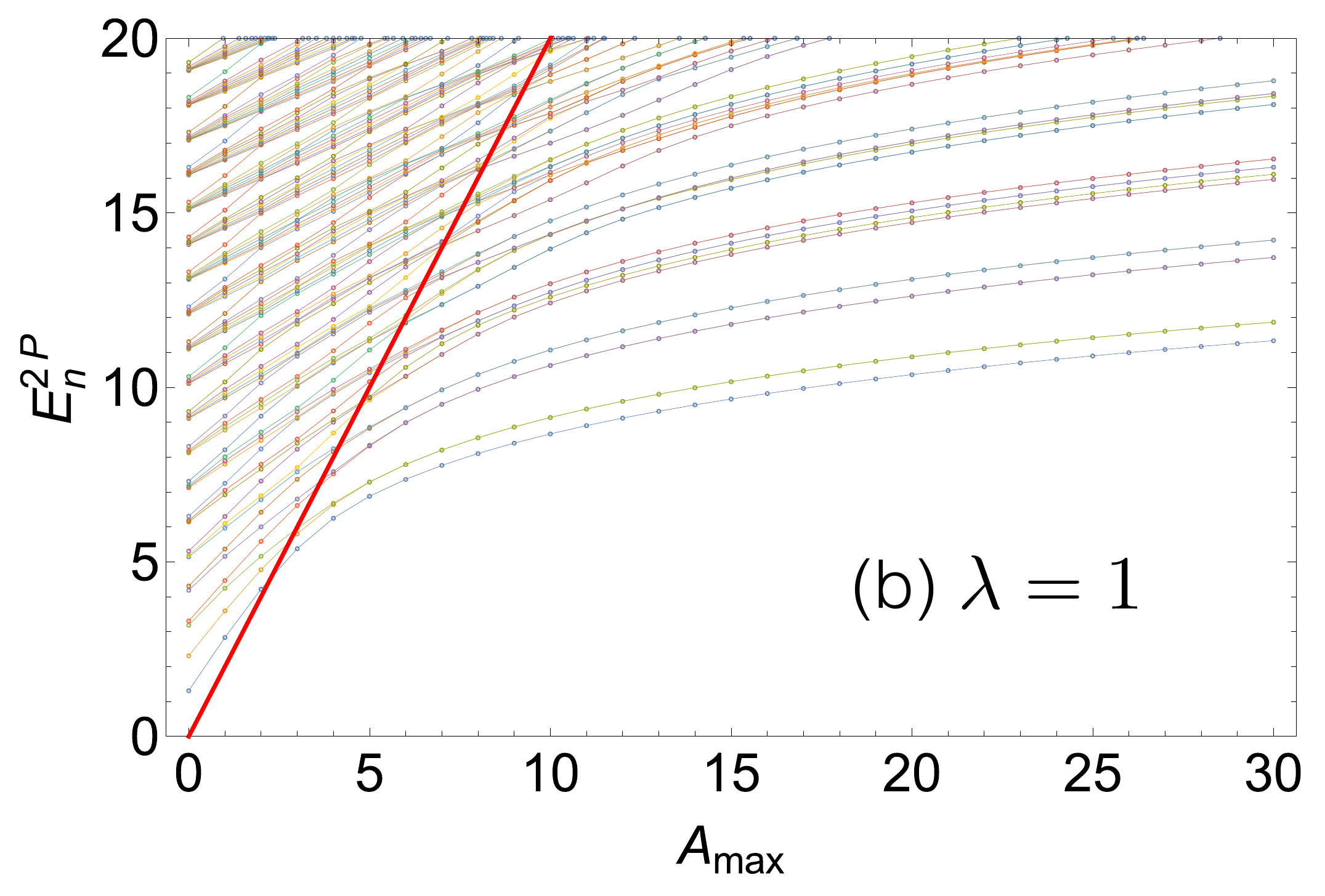}
\caption{(Color online) Two-particle eigenenergies $E_n^{2P}$, Eq.~(\ref{eigenenergies_Nparticles}), 
as a function 
of the (static) tunneling barrier height $A_{\rm max}$, for interaction strengths (a) $\lambda=0$ and (b) $\lambda=1$. 
Finite interactions partially or totally lift the degeneracy of the eigenenergies, depending on the considered quantum number.
The red line indicates the effective potential barrier height -- which is twice the barrier height for a single particle, i.e., $2A_{\rm max }$.
Parameter values for the FGH method: $x_{\rm max}=-x_{\rm min}=40$, $N_{\rm cut}=330$, and $N_{\rm Grid}=2047$.
}
 \label{Figure2}   
\end{figure*}

Turning on the interaction changes the structure
of the energy spectrum, as shown in Fig.~\ref{Figure2} (b).
The calculation of the energy spectrum in the general case $A_{\max}\neq0$ requires a numerical treatment, 
whereas an analytical solution exists  for the harmonic trap with $A_{\max}=0$ and $N=2$ \cite{Busch_Wilkens_1998, Sowinski_Rzazewski_2010}.
The most striking feature is the opening of an energy gap, clearly observed at large $A_{\max}$:
At the ground-state level, the  threefold degenerate states for $\lambda=0$ split into a unique ground state which remains unperturbed by the interaction,
plus two (nearly) degenerate excited states which are affected by the non-vanishing interaction strength $\lambda\neq0$.
This behavior was already discussed in Ref.~\cite{Murphy_McCann_2008} for a polynomial double-well.
Our present results show that this effect is also observed in the excitation spectrum below the separatrix  $2A_{\max}$.
For instance, the first excited state manifold of the $\lambda=0$ limit (see Fig.~\ref{Figure2} (a), in the range $A_{\max} \geq 10$), which is fourfold degenerate, splits  (for $\lambda =1$, Fig.~\ref{Figure2} (b)) into two (nearly) degenerate states unperturbed by the interaction,
plus two (nearly) degenerate states slightly shifted by the interaction.
The presence of these energy gaps in the spectrum will be essential for our understanding of the many-particle dynamics discussed in the next sections.

\begin{figure}[t]
\includegraphics[width=\figwidthsingle \columnwidth]{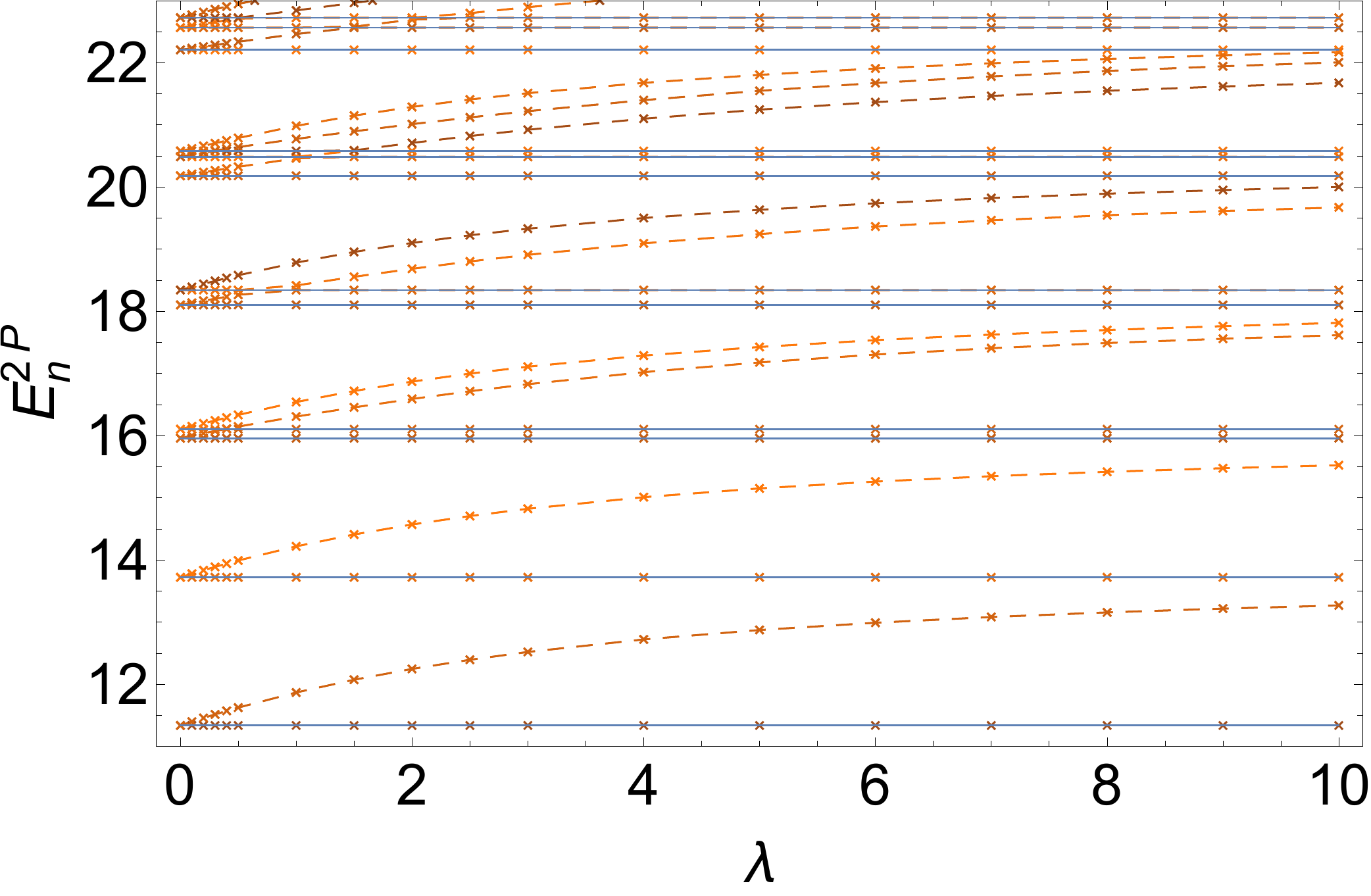}
\caption{ (Color online) Two-particle eigenenergies $E^{2P}_n$, Eq.~(\ref{eigenenergies_Nparticles}), 
as a function of the inter-particle interaction strength $\lambda$, 
in a deep double-well with $A_{\rm max}=30$.
Flat energies (continuous lines) correspond to the situation where the particles are almost completely localized in opposite wells and do not interact. 
Increasing $\lambda$ tends to induce a degeneracy between even and odd states (fermionization process).
FGH parameters: $x_{\rm max}=-x_{\rm min}=40$, $N_{\rm cut}=330$, and $N_{\rm Grid}=2047$.}
\label{Figure3}
\end{figure}

Consideration of a deep double-well, e.g.,  $A_{\rm max}=30$,  allows for a better understanding of interaction-induced spectral features,
as shown in Fig.~\ref{Figure3}.
Indeed, for energies $E^{2P}_n \ll 2A_{\rm max}$, one can approximate the two wells by two decoupled harmonic traps with vanishing tunneling coupling.
Flat energy levels correspond to the situation where the particles are almost completely localized in opposite wells and, consequently, do not interact. 
The remaining energy levels represent configurations where both particles occupy the same well.
The spectral lines then approach the next higher-lying manifold at strong interaction, e.g., $\lambda \simeq 10$. 
In the limit of $\lambda \to \infty$, one recovers the Tonks-Girardeau (or fermionization) limit where these states become degenerate \cite{Murphy_McCann_2008, zollner_few_bosons, zollner_Schmelcher_PRA_2008} with the second excited state manifold. Note that, by construction, this limit
is out of reach for the single-band (or two-mode) approximation widely used in the literature.
Figure~\ref{Figure3} shows that the trend towards degeneracy 
between even and odd states with increasing $\lambda$ (fermionization process) is not restricted to the first spectral manifolds, 
but clearly manifests itself in the entire spectral range $E^{2P}_n \ll 2A_{\rm max}$.
\begin{figure}[h]
	\includegraphics[width=\figwidthsingle \columnwidth]{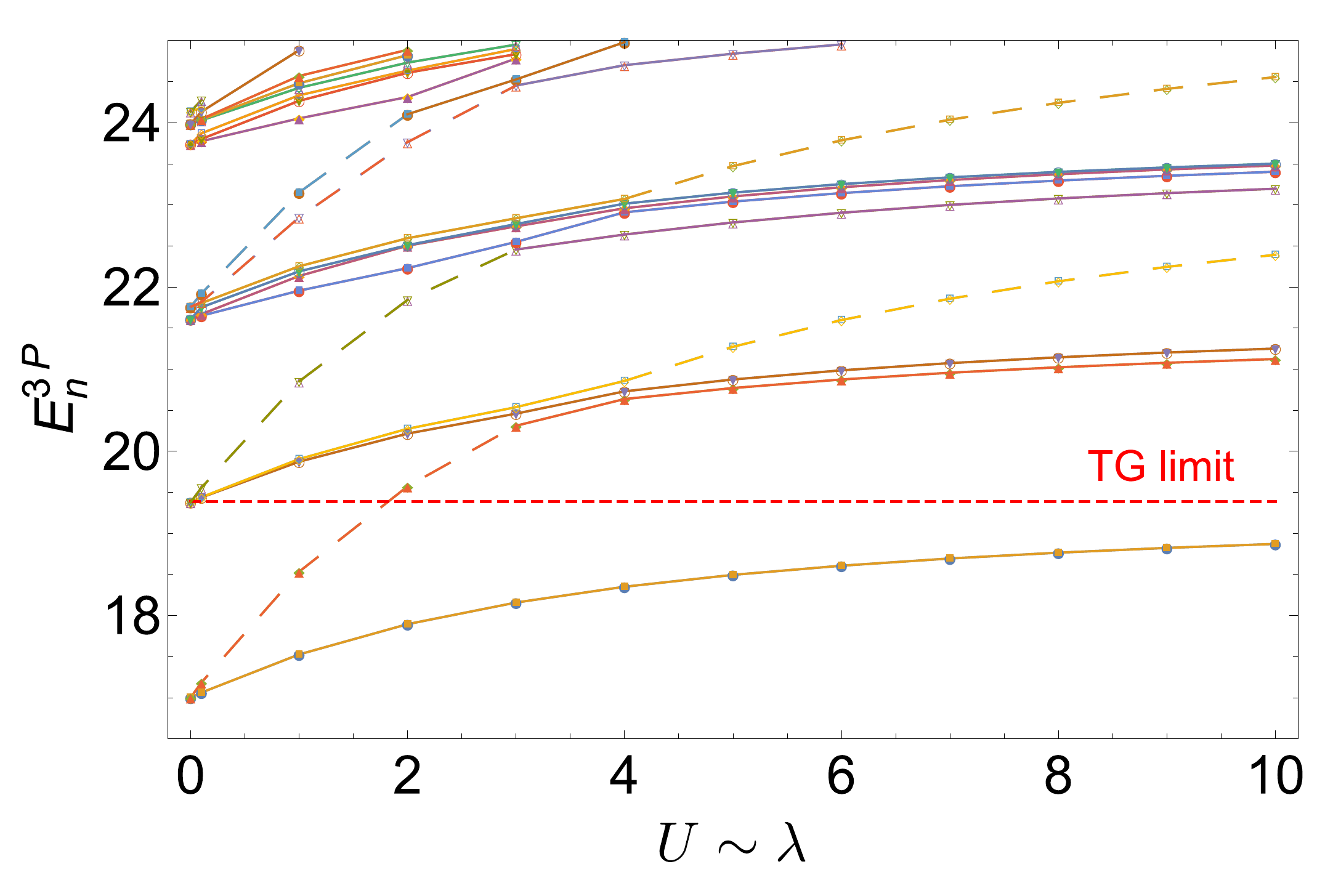}
	\caption{ (Color online) Three-particle eigenenergies $E^{3P}_n$, Eq.~(\ref{eigenenergies_Nparticles}), 
		as a function of the inter-particle interaction 
		strength $U \equiv \lambda  \sum_i  \ | w_{0i} |^4$ (see Appendix~\ref{Method_BH}), with $A_{\rm max}=30$.
		Dashed (continuous) lines represent eigenstates with  three (two) particles on the same well, and 
		the red horizontal line  indicates the Tonks-Girardeau (TG)  limit for the ground state. 
		Parameters employed for the BH method (see App.~B): $x_{\rm max}=-x_{\rm min}=10$, and $L=231$.
	}
	\label{Figure4}
\end{figure}

The situation is (again) very different for three interacting particles \cite{Zollner_Meyer_Schmelcher_PRA_2007}: Figure~\ref{Figure4} shows  the three-particle energy levels, for $A_{\rm max}=30$, as a function  of the interaction strength $U$. All states are sensitive to the interaction and we observe two manifolds of states -- states which exhibit  interactions of two particles (full lines), 
and states which exhibit interactions of three particles (dashed lines).
In contrast to the two-particle case, the ground state remains twofold quasi-degenerate at large  $\lambda$. Note that the present three-particle results were obtained with the BH method (see Appendix~\ref{Method_BH}), 
since the Hamiltonian matrix is sparse in the BH representation, and therefore 
allows for computationally more efficient handling than the FGH method, for which the 
eigenenergies converge only slowly as a function of $N_{\rm cut}$ \cite{FS_thesis_2018}. Furthermore,
in the BH method $U \equiv \lambda  \sum_i  \ | w_{0i} |^4$, cf. Eq.~\eqref{eq:Uvslambda-BH}, substitutes for $\lambda$ used in the FGH calculations.

\subsection{Eigenstate structure and few-body correlations}
\label{sec3_subB}

\begin{figure*}[t]
	\centering
	\includegraphics[width=\figwidthdouble \columnwidth ]{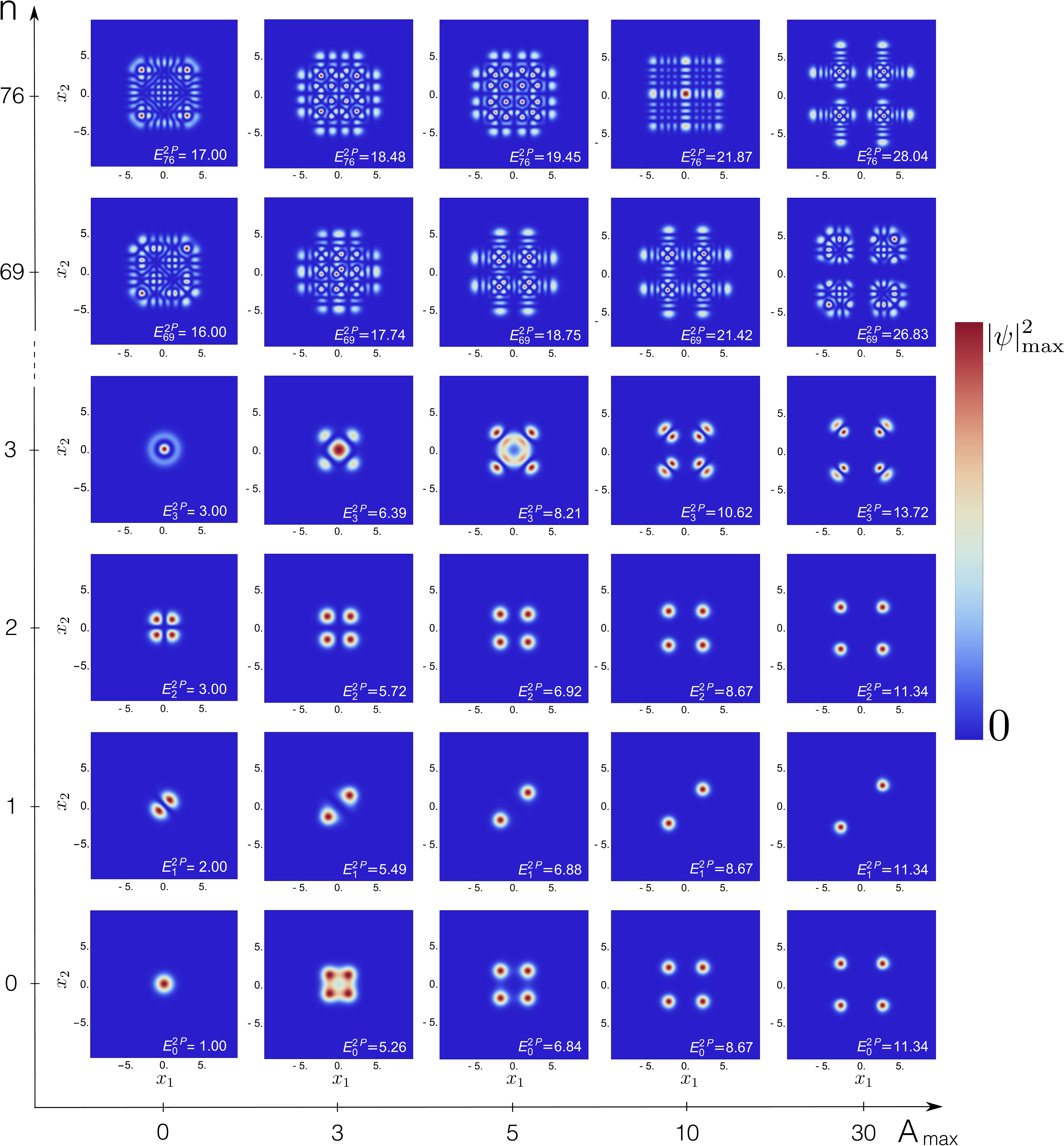}
	\caption{(Color online) Probability densities  $\left|\psi_n(x_1,x_2)\right|^2$ of the $n$th eigenstates of two non-interacting particles ($\lambda$=0), in configuration space $(x_1,x_2)$, with variable barrier height
		from the single ($A_{\rm max}=0$) to the double-well ($A_{\rm max}\neq0$)  scenario, cf. Eq.~\eqref{Double_Well_Potential}. The densities are plotted on a linear scale which interpolates between vanishing probability (dark blue) and the maximum probability density  $\left|\psi\right|_{\rm max}^2$ of the given eigenstate.
		FGH parameters: $x_{\rm max}=-x_{\rm min}=40$, $N_{\rm cut}=330$, and $N_{\rm Grid}=2047$.}
	\label{Figure5}
\end{figure*}

Let us now inspect the associated many-particle eigenstates and the spatial correlations encoded into them, again as a function of both the central barrier height $A_{\rm max}$
and the interaction strength $\lambda$.
The  probability density, Eq.~\eqref{probability_densities}, provides useful intuition.
For two non-interacting bosons, the probability densities $\left|\psi_n(x_1,x_2) \right|^2$ are plotted in Fig.~\ref{Figure5},
for energetically low- and high-lying eigenstates, as well as for different choices of the barrier height $A_{\rm max}$.

At low energies $(n=0,1,2)$, and with increasing barrier height $A_{\rm max}\rightarrow\infty$, $\left|\psi_n(x_1=0,x_2) \right|^2 \to 0$ and   $\left|\psi_n(x_1,x_2=0) \right|^2 \to 0$. 
Consequently, the maxima  of the probability density symmetrically split into the two or four corners of configuration space \cite{Murphy_Busch_2007, Murphy_McCann_2008, Sakmann_Cederbaum_PRA_2008}. For $n=1$, the nodal line $x_1=-x_2$ originates from the superposition of even and odd (nearly) degenerate single-particle states.
Note that the associated eigenenergies are quasi-degenerate at $A_{\rm max}=30$: $E^{2P}_{n=0, 1, 2}\simeq 11.34$.
At higher excitations, where the spectrum must progressively approach that of a harmonic oscillator [recall Fig.~\ref{Figure1}(b)], the eigenstates exhibit a metamorphosis, sometimes even displaying a maximum at the saddle-point, see, e.g., $n=76$, $A_{\rm max}=10$, and thus reminiscent of barrier states of the single-particle problem.

\begin{figure*}[t]
	\centering
	\includegraphics[width=\figwidthdouble \columnwidth ]{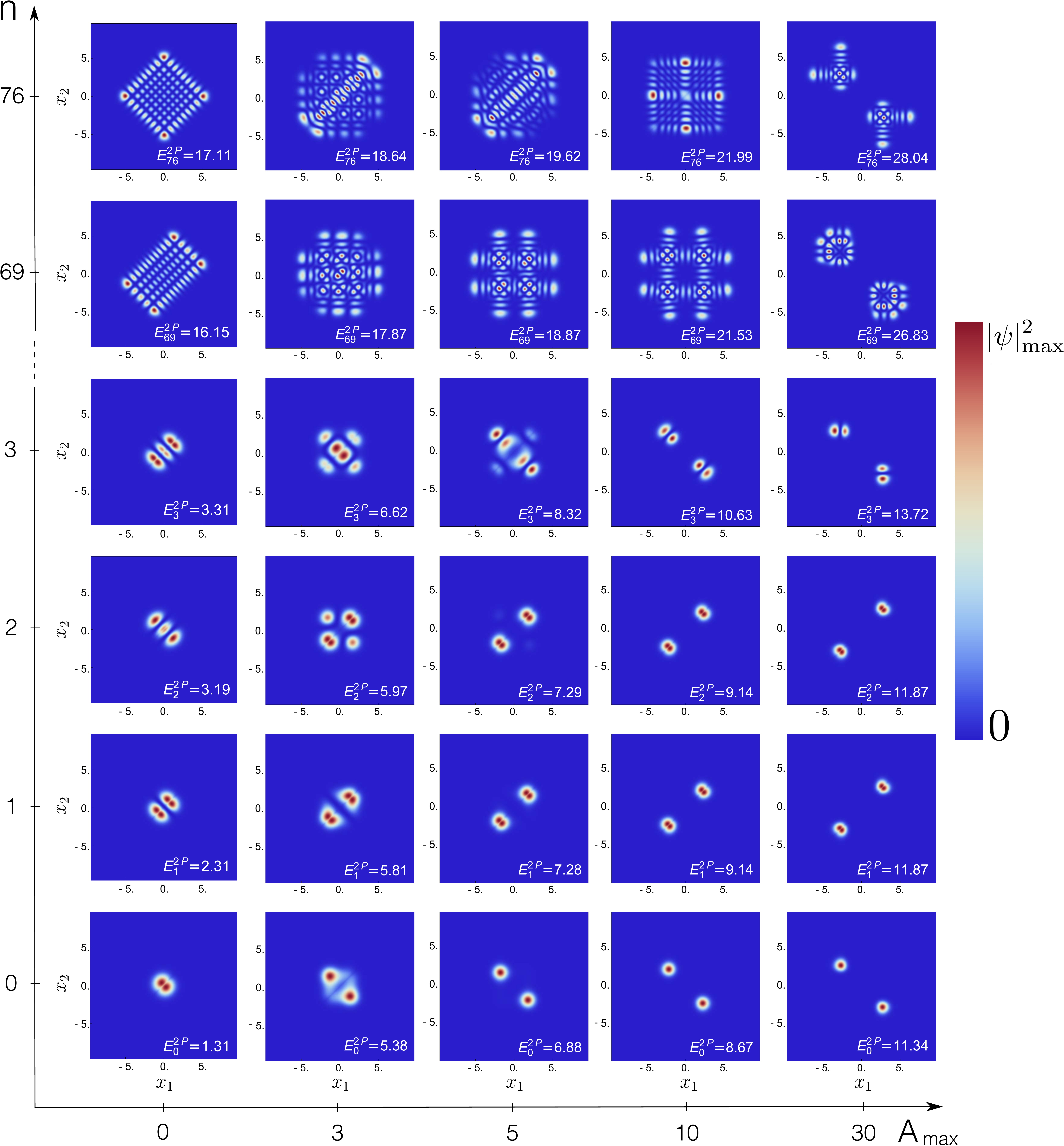}
	\caption{(Color online)  Probability densities  $\left|\psi_n(x_1,x_2)\right|^2$ of the $n$th eigenstates of two interacting particles ($\lambda$=1), in configuration space $(x_1,x_2)$, with variable barrier height
		from the single ($A_{\rm max}=0$) to the double-well ($A_{\rm max}\neq0$)  scenario, cf. Eq.~\eqref{Double_Well_Potential}. Color coding as in Fig. \ref{Figure5}.
		FGH parameters: $x_{\rm max}=-x_{\rm min}=40$, $N_{\rm cut}=330$, and $N_{\rm Grid}=2047$.}
	\label{Figure6}
\end{figure*}
Interactions affect the spatial correlations in many ways, as 
shown in  Fig.~\ref{Figure6} for $\lambda=1$:
Comparison to Fig.~\ref{Figure5} shows that for $n=0 - 3$, the interaction slightly stretches the maxima of the eigenstates along the anti-diagonal  $x_2=-x_1$ \cite{Murphy_McCann_2008}, and in some cases suppresses the amplitudes for double-occupancy of either site or that of delocalization over both sites.
In a deep double-well, e.g., $A_{\rm max}=30$, the  threefold (nearly) degenerate non-interacting eigenstates $n=0-2$ of Fig.~\ref{Figure5}
split into a unique  ground state state and two (nearly) degenerate eigenstates  $n=1, 2$.
At higher excitations ($n=76$), we observe product states in the relative $\propto x_1-x_2$ and center-of-mass $\propto x_1+x_2$  coordinates
(see Fig.~\ref{Figure6} 
for $A_{\rm max}=0$), and, therefore, also for these states correlated tunnelling is expected, as opposed to the independent tunnelling imprinted into the eigenstates in Fig. \ref{Figure5}. The impact of interactions on states in the vicinity of the separatrix is mainly highlighted by a suppression of the density maximum around $x_1=x_2=0$, see  the result for $A_{\rm max}=10, n=76$ in Fig.~\ref{Figure6}.

\begin{figure*}[t]
	\includegraphics[width=2.0\columnwidth]{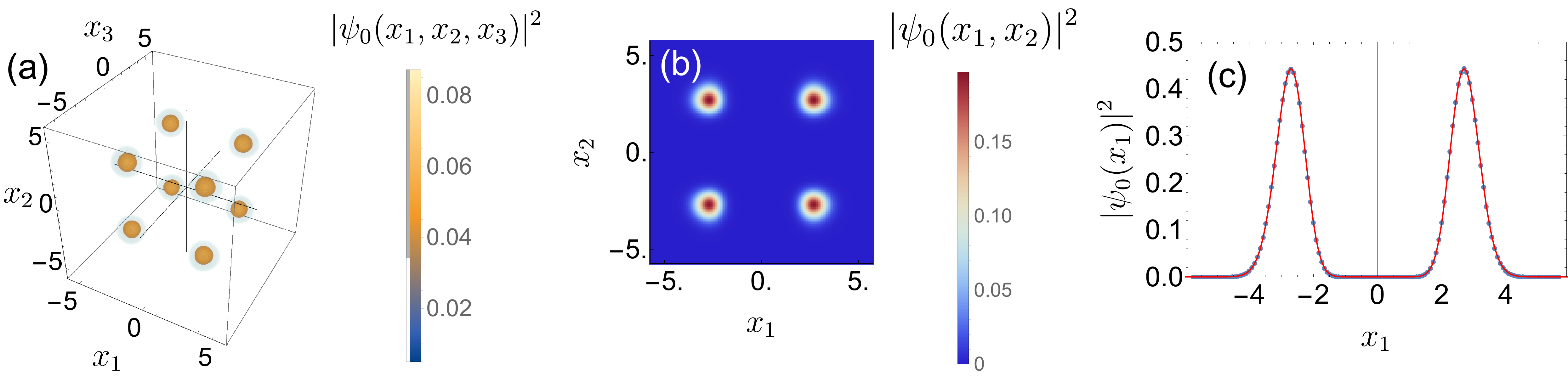}\\
	\includegraphics[width=2.0 \columnwidth]{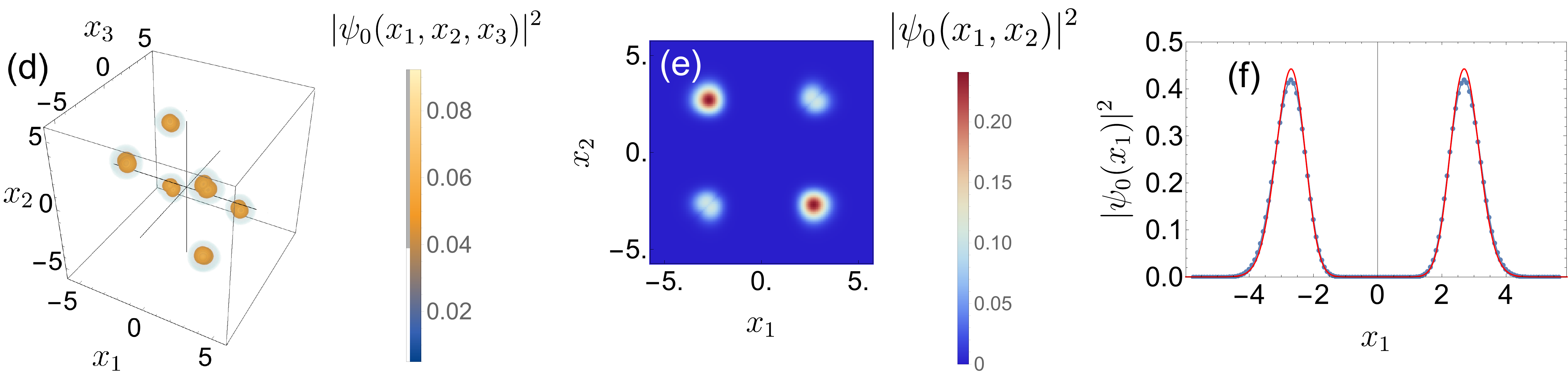}
	\caption{(Color online) 
		Three-body probability density  $\left|\psi_0(x_1,x_2,x_3)\right|^2$ (a,d), diagonal of the reduced two-body probability density matrix $\left|\psi_0(x_1,x_2)\right|^2$ (b,e), and diagonal of the reduced one-body probability density matrix $\left|\psi_0(x_1)\right|^2$ (c,f) of the ground state of three (a--c) non-interacting ($U=0$) and  (d--f) interacting ($U=1$) particles in the double well ($A_{\rm max}=30$), cf. Eq.~\eqref{Double_Well_Potential}.
		Note that, in (d), $|\psi_0|^2\approx 0$ if all bosons are in the same well ($x_1,x_2,x_3>0$ and $x_1,x_2,x_3<0$), due to the interactions.
		The red line in (c) is the profile of $|\psi_0(x_1)|^2$ for non-interacting particles.
		Parameters employed for the BH method: $x_{\rm max}=-x_{\rm min}=10$, and $L=231$.}
	\label{Figure7}
\end{figure*}

Next, let us have a closer look at the  three-body probability density $\left|\psi_n(x_1,x_2, x_3) \right|^2$ of the ground state $(n=0)$ in a deep double-well, $A_{\rm max}=30$. 	
Figures~\ref{Figure7} (a) and (d) show the 
three-body probability density \eqref{probability_densities} for non-interacting,  $U=0$, and interacting, $U=1$, particles, respectively [see Eq.~\eqref{eq:Uvslambda-BH}].  
Since all particles occupy the same single-particle orbital $ | \psi_0 \rangle $, the non-interacting ground state  covers all eight octants of configuration space in Fig.~\ref{Figure7} (a).

Like in the two boson case, the three-body wave function develops a nodal line along the  main diagonal  $x_1=x_2=x_3$ for non-vanishing $U>0$. At strong interaction, the maxima of the wave function are additionally shifted towards the corners of configuration space, along the diagonals $x_1=x_2=-x_3$, $x_1=-x_2=x_3$ and $-x_1=x_2=x_3$.
Using a two-mode description, the ground state for sufficiently strong interactions is given by two particles at the same site and one on the opposite site. Therefore, the ground state, illustrated in Fig.~\ref{Figure7} (d), has no density in the areas associated with three particles at the same site ($x_1,x_2,x_3>0$ and $x_1,x_2,x_3<0$). 
Moreover, the two mode description in the Fock basis $\ket{n_L,n_R}$ helps to understand the structure of the doubly degenerate ground state, 
since both states
\begin{align}
\ket{\psi_1}&= \ket{2,1}, \notag \\
\ket{\psi_2}&= \ket{1,2},
\end{align}
give rise to the same energy.
The degenerate first and second excited states are then given by 
\begin{align}
\ket{\psi_3}&= \ket{3,0}, \notag \\
\ket{\psi_4}&= \ket{0,3},
\end{align}
which are strongly sensitive to the interaction. Therefore, the fourfold degenerate ground state in the non-interacting case evolves into two doublets of states which further separate as a function of the interaction strength, as illustrated in the spectrum in Fig.~\ref{Figure4}.

Finally, we inspect how the correlation information imprinted into the three-particle state is reduced when subsequently integrating out degrees of freedom. 
Averaging over one degree of freedom leads to the diagonal of the reduced two-body density matrix
$|\psi_0(x_1, x_2)|^2= \int \text{d}x_3 |\psi_0(x_1, x_2, x_3)|^2$,
plotted for $U=0$ and for $U=1$ 
in Figs.~\ref{Figure7} (b) and (e), respectively. 
The impact of interaction becomes clearly visible by the reduction of the density 
along the diagonal $x_1=x_2$, tantamount of reduced correlations -- as already observed 
in Figs.~\ref{Figure5} and \ref{Figure6}. 
Note that, in some contrast to the density of the two-particle state $n=0$,
for $\lambda$=1 and $A_{\rm max}=30$ in Fig.~\ref{Figure6}, the 
probability to detect two particles in the same well  is not fully suppressed 
at interaction strength $U=1$.

Averaging over the second degree of freedom leads to the diagonal of the reduced one-body density matrices, $|\psi_0(x_1)|^2= \int \text{d}x_2 \text{d}x_3 |\psi_0(x_1, x_2, x_3)|^2$,  displayed in Figs.~\ref{Figure7} (c) and (f). 
The profile of $|\psi_0(x_1)|^2$ for $U=0$,  cf. Fig.~\ref{Figure7} (c), is exactly the same as the one obtained for the non-interacting two-particle case (red line), as expected. 
Only a small difference between the one-body densities $|\psi_0(x_1)|^2$  associated with interacting  and non-interacting (red line) bosons, respectively, is detectable, cf. Fig.~\ref{Figure7} (f) \footnote{Note that the two-mode approximation (i.e., the double well Bose Hubbard model) is not sensitive to changes of the intra-well correlations -- which here  manifest themselves in the changed one-body density profile.}.
This analysis therefore indicates that even if the interaction strongly affects the correlations, this information is not reflected by the one-body density profile.  

\FloatBarrier

\section{Dynamics in the double well}
\label{sec4}
\subsection{Static potential: two-body excited state dynamics}
\label{sec4_subA}

Given the above phenomenology of spectra and eigenstates, we now explore how the tunneling dynamics of two interacting particles in a  static double-well depends on the choice of the initial state.
To this end, we  consider a system initially prepared in a (non-stationary) superposition of excited states, 
such that both particles are localized on the right-hand side of the double-well, 
at fixed barrier height $A_{\rm max}=10$.
This localized state can be constructed  by coherent superposition of (non-interacting) adjacent, even and odd one-body eigenstates:
\begin{equation}
\ket{\Psi_n^{\rm loc}(t=0)}= \frac{1}{2} \left(\ket{\Psi^{1P}_{2n+1}}+\ket{\Psi^{1P}_{2n}}\right) \otimes  \left(\ket{\Psi^{1P}_{2n+1}}+\ket{\Psi^{1P}_{2n}}\right) .
\label{2particle_excited_intial_state}
\end{equation}
The dynamics is deduced from a spectral decomposition of the 
many-body Hamiltonian \eqref{MB_Hamiltonian} with the  FGH method, and we compare
the dynamics seeded by a low-lying initial state $\ket{\Psi^{\rm loc}_{n=0}(t=0)}$  
to that of an initial state $\ket{\Psi^{\rm loc}_{n=3}(t=0)}$  with energy close to the potential's saddle-point, i.e., $E^{2P}\simeq 20$, see Fig. \ref{Figure1}(a).

In the non-interacting case, the wave function remains separable at all times and, therefore, one can straightforwardly 
express the probabilities \eqref{eq:detection-probabilities} in terms of the single-particle density, which yields
\begin{align}
\nonumber
P_{(LL)}(t) &= P^2_L(t) =  \left[ \int_{x_{\rm min}}^{0} \text{d}x ~ |\psi(x;t)|^2 \right ]^2  ,\\
\label{eq:detection-probabilities_bis}
P_{(LR)}(t) &=  2 \cdot P_L(t)P_R(t),\\
\nonumber
P_{(RR)}(t)&= P^2_R(t) =  \left[ \int_{0}^{x_{\rm max}} \text{d}x ~ |\psi(x;t)|^2 \right ]^2   .
\end{align} 
Applying a simplified three-level model for $n=0$ \cite{sh_diss},
Eq.~(\ref{eq:detection-probabilities_bis}) can be rewritten as
\begin{align}
P_{(LL)}(t) &= \sin^4\left(\frac{\Delta}{2} t \right) ,\notag \\
P_{(LR)}(t) &= \frac{1}{2}\sin^2\left(\Delta t\right),\\
P_{(RR)}(t) &=  \cos^4\left(\frac{\Delta}{2} t \right). \notag
\end{align} 
Due to the equidistance between the low-lying energies $E^{2P}_{2},  E^{2P}_{1},$ and  $E^{2P}_{0}$ for $\lambda=0$, 
the uncorrelated tunneling dynamics is governed by a single Rabi frequency 
$\Delta= E^{2P}_{2}- E^{2P}_{1}=E^{2P}_{1}- E^{2P}_{0}$ \cite{zollner_few_bosons, zollner_Schmelcher_PRA_2008}.
In particular,  for $A_{\rm max}=10$, $P_{(LL)}(t)$ and $P_{(RR)}(t)$  oscillate with the period  $T(\lambda=0)=2\pi/\Delta \simeq 12\cdot 10^3$.

A finite interaction strength perturbs the equidistance between the low-lying energies and, therefore, two distinct periods 
emerge from the dynamics: $T_{21}= 2\pi/(E^{2P}_{2}- E^{2P}_{1})$ and $T_{10}=2\pi/(E^{2P}_{1}- E^{2P}_{0})$, 
in qualitative agreement with experimental observations \cite{Folling_Bloch_Nature_2007}.
\begin{figure}[t]
\includegraphics[width=1 \columnwidth]{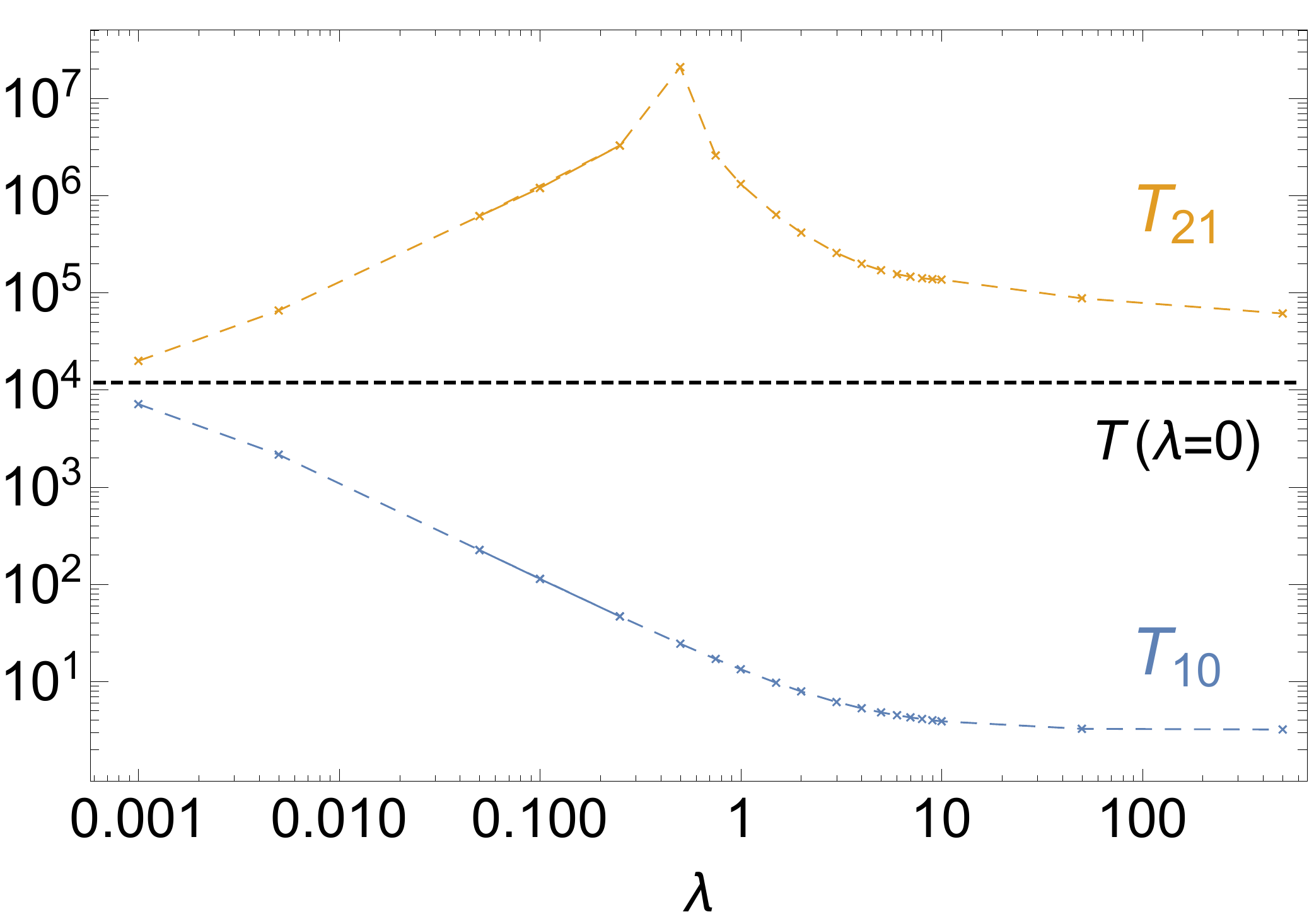}
\caption{(Color online) 
Characteristic periods $T_{21}$ and $T_{10}$ of the two-particle tunneling dynamics as displayed in Fig.~\ref{Figure9}, as a function of the interaction strength $\lambda$, for 
a double-well potential barrier height $A_{\rm max}=10$, 
on a double-logarithmic scale. 
The horizontal, black, dashed line indicates the (degenerate, see main text) period of the non-interacting case $T(\lambda=0)\simeq12\cdot10^3$.}
 \label{Figure8}   
\end{figure}
\begin{figure*}[t]
\includegraphics[width=\figwidthsingle \columnwidth]{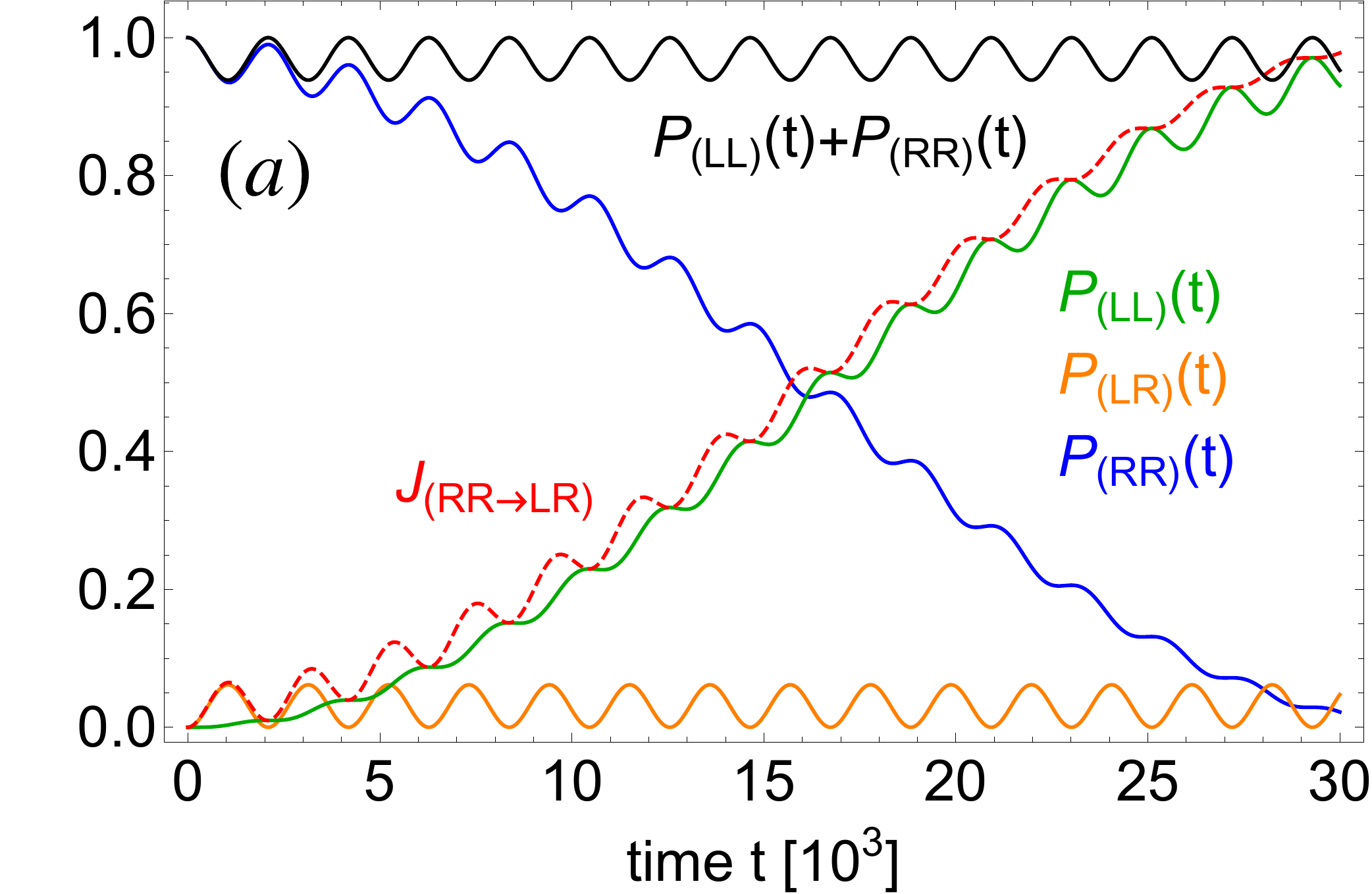}
\includegraphics[width=\figwidthsingle \columnwidth]{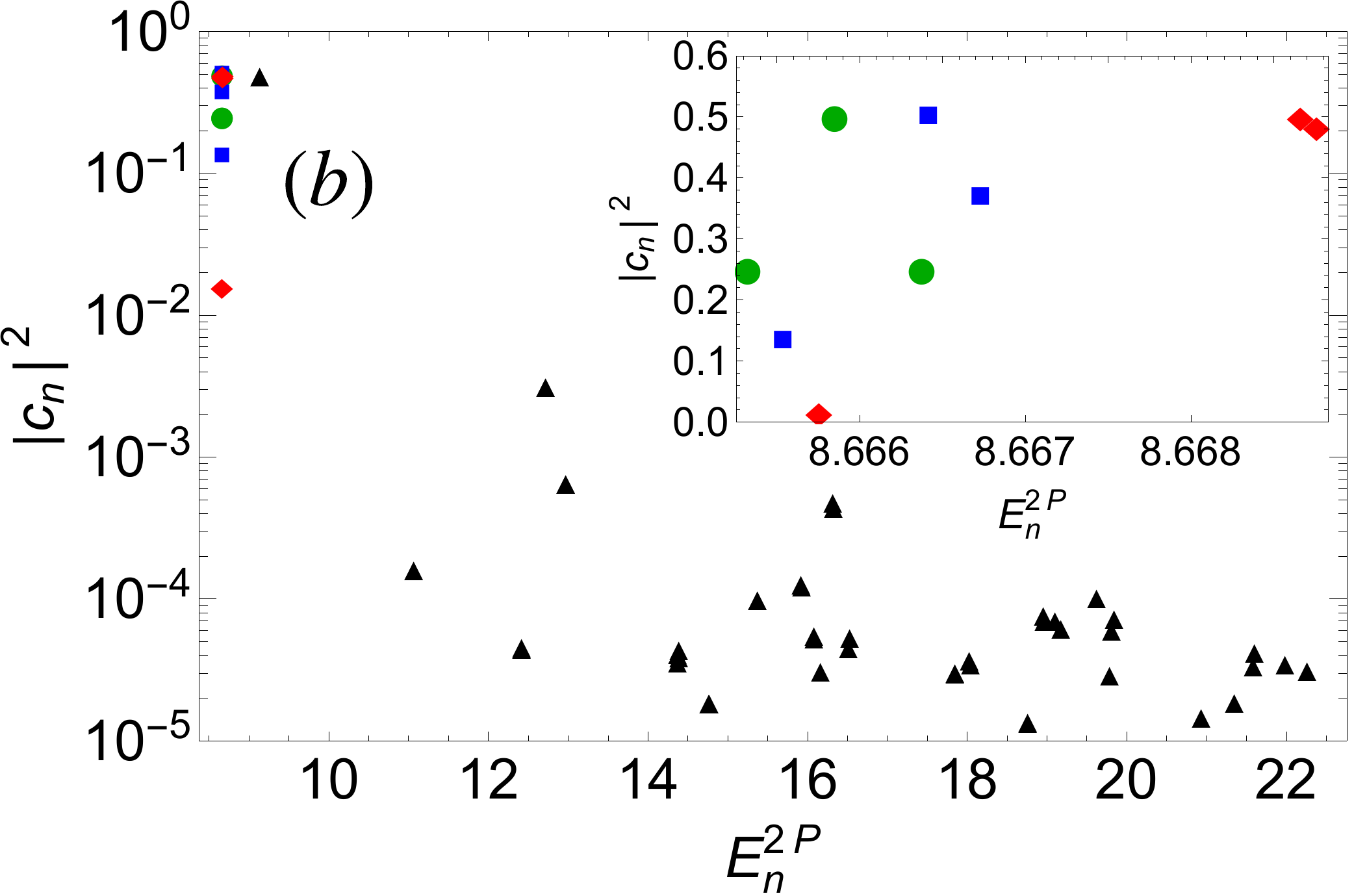}
\caption{(Color online) 
(a) Detection probabilities, Eq.~\eqref{eq:detection-probabilities}, and  time integrated probability current,
Eq.~\eqref{eq:josephson-probabilityflux}, as a function of time,
for the two particle initial state $\ket{\psi^{\rm loc}_{n=0}(t=0)}$, Eq.~(\ref{2particle_excited_intial_state}), and  a weak interaction strength  $\lambda=0.005$.  
(b) Expansion coefficients of the initial state in the interacting two-body eigenbasis, as a function of the  eigenenergy $E^{2P}_n$, 
for interactions  $\lambda=0 ~(\text{circles}),~ 0.001~ (\text{squares}),~ 0.005~ (\text{diamonds})$ and $1~ (\text{triangles})$.
The inset zooms onto the dominant expansion coefficients. FGH parameters: $x_{\rm max}=-x_{\rm min}=40$, $N_{\rm cut}=330$, and $N_{\rm Grid}=2047$.
}
 \label{Figure9}   
\end{figure*}
The evolution of these periods with $\lambda$, plotted in Fig.~\ref{Figure8}, shows a rapid increase (decrease) of 
$T_{21}$ ($T_{10}$) for weak interactions $\lambda<0.5$, and a monotonous decrease of $T_{21}$ for $\lambda>0.5$, while $T_{10}$ saturates at $T_{10}\simeq3$ for $\lambda \to \infty$.
Note that for $\lambda=0.5$, the Josephson oscillation period $T_{21} \sim 1750 \cdot T(\lambda=0)$ is much larger than the 
one for non-interacting particles -- but \textit{finite}.
This corresponds to the self-trapping regime \cite{Albiez_Oberthaler_PRL_2005}.
Interestingly, the Josephson oscillation period $T_{21}$ converges  to the non-interacting  period, $T_{21}\sim T(\lambda=0)$, 
in the Tonks-Girardeau limit $\lambda \to \infty$.
This effect is in agreement with the fermionized pair-state dynamics  discussed in Refs.~\cite{zollner_few_bosons, zollner_Schmelcher_PRA_2008}.

In the two-mode approximation (i.e., the double-well Bose-Hubbard model) for the present scenario, the dynamics is fully described by the amplitudes of the Fock basis states $|n_L, n_R\rangle \in \{|2, 0\rangle,|1, 1\rangle, |0, 2\rangle \}$, with degenerate $|2, 0\rangle$ and  $|0, 2\rangle$.
Two correlated two-particle tunneling processes are then possible in this simplified picture:
a first-order, two-particle tunneling process which corresponds to the direct tunneling of both bosons along the diagonal $x_1=x_2$ 
(i.e.,  the transition $|2, 0\rangle \to |0, 2\rangle$), or a second-order process (i.e., the  transition $|2, 0\rangle \to |1, 1\rangle \to |0, 2\rangle$).
We now elucidate the actual nature of the tunneling process, for weak interactions.

Starting in the initial state $\ket{\Psi^{\rm loc}_{n=0}(t=0)}$ as defined by \eqref{2particle_excited_intial_state}, with  $\lambda=0.005$, 
the dynamics clearly exhibits the Josephson oscillation period $T_{21}\simeq 65 \cdot 10^3$, garnished by a small amplitude beat frequency associated with $T_{10}\simeq 2 \cdot 10^3$. 
These oscillations are observed in the time evolution of the detection probabilities \eqref{eq:detection-probabilities} in 
Fig.~\ref{Figure9} (a), with the Josephson oscillation period  $T_{21}\simeq 5.5  \cdot T(\lambda=0)$ strongly enhanced with respect to the non-interacting value $T(\lambda=0)$. This is in good qualitative agreement with  
experimental observation \cite{Folling_Bloch_Nature_2007}.
One also encounters a strongly reduced probability to 
observe the bosons in opposite wells, signaled by ${\rm max} (P_{(LR)})<0.1$ in Fig.~\ref{Figure9} (a).
The reduction of  ${\rm max} (P_{(LR)})$, arising from the interaction between the particles, 
suggests a direct tunneling along the diagonal $x_1=x_2$, i.e.,  a first-order  tunneling process.
Such a reduction, which is a corollary of
$P_{2}(t) \equiv \int_{x_1\cdot x_2 \geq0} \text{d}x_1  \text{d}x_2 ~ |\psi(x_1,x_2;t)|^2 =P_{(LL)}(t) +P_{(RR)}(t) =1-P_{(LR)}(t) \lessapprox 1$,
was previously discussed in  Refs.~\cite{zollner_few_bosons, zollner_Schmelcher_PRA_2008}.
However, its interpretation as evidence of first-order tunneling is in contradiction with the time dependence of the integrated probability current 
$J_{(RR\to LR)}(t)$ also shown in Fig.~\ref{Figure9} (a), which  clearly indicates a transport across the domain $(LR)$.
Indeed, $J_{(RR\to LR)}(t)$  records all probability which passes $(LR)$ 
and \textit{excludes} the tunneling along the diagonal $x_1=x_2$. This quantity thus allows us
to discriminate sharply the two types of two-particle tunneling.
By virtue of Fig.~\ref{Figure9} (a), 
$J_{(RR\to LR)}(t)\sim P_{(LL)}(t)$ implies that
almost all probability that oscillates between regions $(LL)$ and $(RR)$ passes region $(LR)$. 
This confirms a second-order rather than direct first-order  tunneling from region $(LL)$ to $(RR)$.

\begin{figure*}[t]
\includegraphics[width=\figwidthsingle \columnwidth]{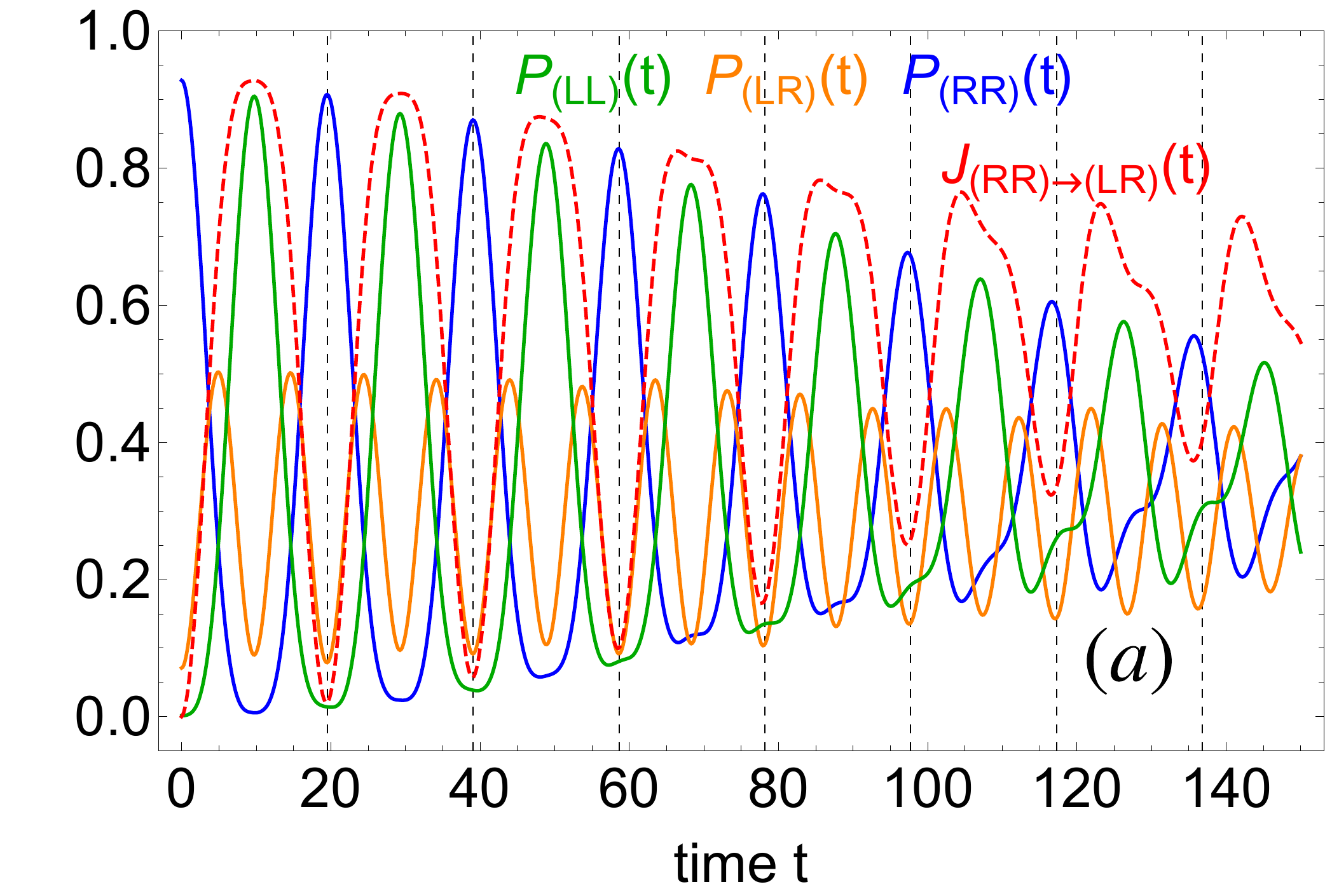}
\includegraphics[width=\figwidthsingle \columnwidth]{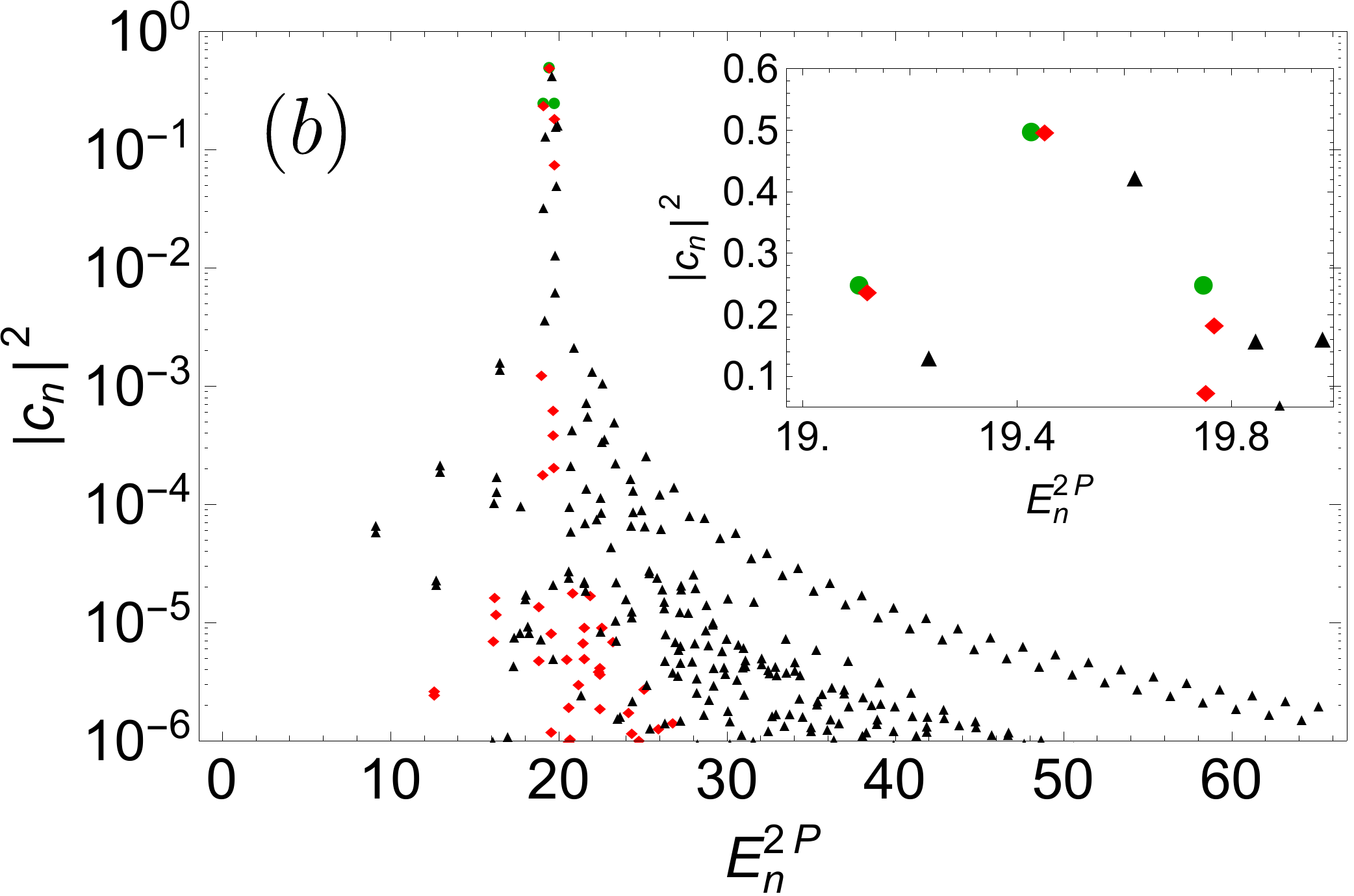}
\caption{
(Color online) 
(a) Detection probabilities \eqref{eq:detection-probabilities}, and  time integrated probability current 
\eqref{eq:josephson-probabilityflux}, as a function of time,
with initial two particle state $\ket{\Psi^{\rm loc}_{n=3}(t=0)}$, Eq.~(\ref{2particle_excited_intial_state}), and interaction strength $\lambda=0.1$. The vertical, dashed black lines indicate the period $T(\lambda=0)\simeq19.5$ of the non-interacting case.
(b) Expansion coefficients of the initial state in the interacting two-body eigenbasis, as a 
function of the  eigenenergy $E^{2P}_n$,  for interaction strengths $\lambda=0 ~(\text{circles}),~ 0.1~ (\text{diamonds})$, and $1~ (\text{triangles})$.
The inset zooms onto the dominant expansion coefficients. FGH parameters: $x_{\rm max}=-x_{\rm min}=40$, $N_{\rm cut}=330$, and $N_{\rm Grid}=2047$.
}
 \label{Figure10}   
\end{figure*}

An explanation of the underlying mechanism follows from the expansion coefficients of  $\ket{\Psi^{\rm loc}_{n=0}(t=0)}$ in the interacting two-particle basis. 
The inset in Fig.~\ref{Figure9} (b) shows that,  for non-interacting particles, 
only three coefficients -- associated with equidistant energies -- are non-zero, 
giving rise to the single frequency $\Delta= E^{2P}_{2}- E^{2P}_{1}=E^{2P}_{1}- E^{2P}_{0}$ oscillations described above. 
Turning on a weak interaction [e.g., $\lambda \leq 0.005$, in Fig.~\ref{Figure9}(a)], the initial state's overlap with the ground state decreases, while, 
at the same time, the coefficients of the first two excited states pick up comparable weights (squares and diamonds in the inset).
The mechanism behind the observed tunneling process is straightforward:  
in the previous Section, we showed that the first two excited states 
stick together to form a doublet with an energy which increases with $\lambda$,
while the energy of the ground state -- one particle localized on each well -- does not depend on the interaction, 
cf. Fig.~\ref{Figure3}. 
Therefore, the ground state corresponding to a balanced population in region (LR), see Fig.~\ref{Figure6}, becomes off-resonant. 
Thus, if a boson tunnels from the right- to the left-hand side, it can populate the ground state only for very short times. The associated time-scale is determined by the energy gap between the ground state and the degenerate excited states' energy. 
Subsequently, the boson tunnels either back to the right well, or the other boson tunnels from the right to the left well, to re-establish energy conservation. 
It follows from this latter argument that the involved frequencies can be inferred from a three-level model \cite{sh_diss}. 
Increasing further the interaction, the excited states turn resonant with the next higher-lying band (recall Figs.~\ref{Figure2}(b) and \ref{Figure3}), 
such that additional transitions kick in, and the tunneling dynamics exhibits more frequencies, with no simple representation in the 
above three-level model. In terms of the expansion coefficients, 
this boils down to an increasing number of contributing eigenstates as illustrated, for $\lambda=1$, by the triangles in Fig.~\ref{Figure9} (b). \\

Considering now the non-interacting, excited initial state $\ket{\psi^{\rm loc}_{n=3}(t=0)}$ [see Eq. \eqref{2particle_excited_intial_state}] with energy close to the saddle-point, i.e., $E^{2P}\simeq 20$, 
the  uncorrelated tunneling dynamics (not shown) is that of a separable wave function with a single Rabi frequency 
$\Delta= E^{2P}_{59}- E^{2P}_{52}=E^{2P}_{52}- E^{2P}_{51}$,
and period $T(\lambda=0)= 2 \pi / \Delta \simeq 19.5$. 
This monochromaticity again is a consequence of the  equidistant level spacing of the high-lying energies $E^{2P}_{59},  E^{2P}_{52}$,  and $E^{2P}_{51}$, for $\lambda=0$ [see circles inset Fig~\ref{Figure10} (b)].
Note that the Rabi period $T \simeq 19.5$ is much smaller than the one observed for the initial condition $\ket{\psi^{\rm loc}_{n=0}(t=0)}$, for which $T \simeq 12\cdot 10^3$, since $E^{2P}_{52}- E^{2P}_{51} > E^{2P}_{1}- E^{2P}_{0}$, and the detection probabilities, Eq.~\eqref{eq:detection-probabilities}, oscillate with reduced amplitude (smaller than 1), due to a less pronounced localization of  $\ket{\psi^{\rm loc}_{n=3}(t=0)}$ in either one of the individual wells.

How do interactions affect the evolution of the initial state $\ket{\Psi^{\rm loc}_{n=3}(t=0)}$?
As expected from our above spectral analysis, much stronger interactions than $\lambda=0.005$ must be considered to 
induce visible effects in the dynamics, since the impact of interactions is comparable for all eigenstates (cf. Fig.~\ref{Figure6}, for $A_{\rm max}=30$ and $n=76$) which exhibit 
a large overlap with the initial state. 
Figure~\ref{Figure10} (a) shows 
the time-evolution of the detection probabilities \eqref{eq:detection-probabilities} for $\lambda=0.1$.
The oscillation period seeded by $\ket{\Psi^{\rm loc}_{n=3}(t=0)}$ appears to be much less sensitive to interactions than for $\ket{\Psi^{\rm loc}_{n=0}(t=0)}$ (recall Fig.~\ref{Figure9}):
the oscillation periods  of $P_{(LL)}(t) $ and $P_{(RR)}(t)$ almost coincide with the non-interacting period 
 $T(\lambda=0)  \simeq 19.5$ indicated by vertical dashed lines.
Nevertheless, a small shift is visible after seven periods around $t\simeq136.5$.
This small shift  can be understood by inspection of the
 expansion coefficients of the initial state in the interacting two-body eigenbasis, 
Fig.~\ref{Figure10} (b).
In contrast to $\lambda=0$, where only three energy levels contribute to the dynamics [circles, inset Fig.~\ref{Figure10} (b)], 
an interaction $\lambda=0.1$ redistributes the amplitudes over four dominant states with a weight larger than $5\%$ [squares, inset Fig.~\ref{Figure10} (b)].
The interactions slightly modify the energy gaps, leading to a small modification of the Josephson period, and give finite weight to one additional eigenstate, leading to a modulation of the plotted observables with period $T\simeq 394$. 
This additional modulation of the signal must not be confused with the damping of density oscillations as observed for 
large particle numbers in bosonic Josephson junctions \cite{Sakmann_Cederbaum_PRL_2009,sakmann_2014}.
As indicated by the time-integrated probability current which roughly follows $P_{(LL)}(t)$ 
in Fig.~\ref{Figure10} (a), 
we again witness a second-order tunneling across region (LR), 
instead of direct first-order tunneling along the diagonal $x_1=x_2$.
When further increasing the interaction, see, e.g., the diamonds for $\lambda=1$ 
in Fig.~\ref{Figure10} (b), significantly more states contribute to the time evolution (not shown).
The inter-particle interaction enforces mixing of the dynamics in the reduced single-particle subspace, and, accordingly, increases the single-particle entropy.
 
\begin{figure*}[t]
\begin{center}
	\includegraphics[width=\figwidthdouble\columnwidth]{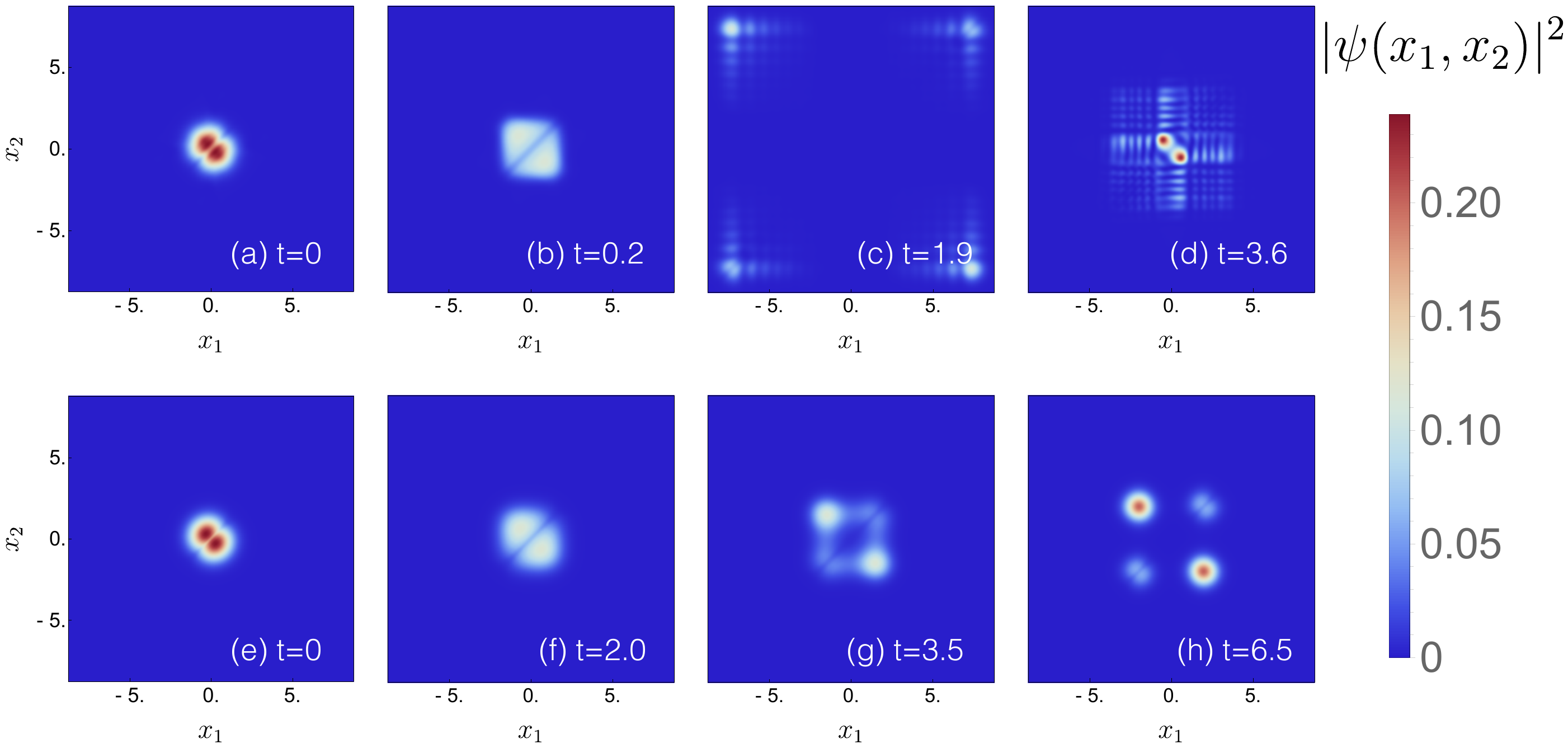}	
	\caption{(Color online) 
	Time evolution of the two-body density $\left|\psi (x_1,x_2;t)\right|^2$,
	launched in the initial two particle ground state of a harmonic trap, with interaction strength $\lambda =1$,
	for (a--d) a diabatically switched central barrier with amplitude $A_{\rm max}=30$
	($T_{\rm ramp} \rightarrow 0$, with FGH parameters $x_{\rm max}=-x_{\rm min}=40$, $N_{\rm cut}=330$, and $N_{\rm Grid}=2047$),
	and (e--h) (quasi-) adiabatic switching to $A_{\rm max}=30$. ($T_{\rm ramp}=30$, with \textsc{Mctdh-x} parameters $x_{\rm max}=-x_{\rm min}=12$, $N_{x}=512$, and $M=20$.)
	}
	\label{Figure11} 
	\end{center}  
\end{figure*}

\subsection{Time-dependent double-well potential: from few- to  many-body dynamics}
\label{sec4_subB}

We have seen in the previous sections how the barrier height affects the impact of interactions on the many-particle dynamics. 
We now generalize this analysis by considering a time-dependent switching of the barrier according to Eqs.~\eqref{Double_Well_Potential} and \eqref{Double_Well_Amplitude}, with $A_{\rm max}=30$. 
Before this quench, the bosons are  prepared in the interacting many-particle ground state of a harmonic trap.
Our purpose is here to examine how the reduced one-body density matrix evolves 
for (quasi)-adiabatic vs. diabatic switching.
Extrapolation to larger particle numbers using the \textsc{Mctdh-x} method relates our observations to previous studies of the splitting of a BEC by a laser sheet \cite{cederbaum_splitting_dynamics, Menotti_2001, Shin_Leanhardt_PRL_2004}.
Note that while quenches can be efficiently simulated with the help of the FGH method, we employ the \textsc{Mctdh-x} method (see Appendix~\ref{Method_MCTDH}) for finite-switching times, to 
deal with the time-dependent Hamiltonian \eqref{MB_Hamiltonian}.

We start with the time evolution of the many-body wave function when 
the tunneling barrier is suddenly quenched from $A_{\rm max}=0$ to 30 (i.e., $T_{\rm ramp} \rightarrow 0$) \cite{Orzel_Kasevich_2001, Mahmud_Reinhardt_2005, Ebert_Hammer_2016}. 
Figure~\ref{Figure11}  (a--d) shows the behavior of the two-particle density for $\lambda=1$, during the initial stage of the quench-induced dynamics.
The initial wave packet is split along the diagonal $x_1 = x_2$,  and  spreads towards the outer edges of the double-well, until its reflection after half a period $t \simeq 1.9$.
Since  all the injected energy, i.e., $A_{\rm max}=30$,  is suddenly transferred to the two bosons, 
the turning point $x_{\rm turn} \sim \pm ~7.75$  in Fig.~\ref{Figure11}  (c),
where the reflection takes place, corresponds to  $V(x_{\rm turn}) \simeq A_{\rm max}=30$ (see Fig.~\ref{Figure1} (b)).
We observe (not shown) that  the higher the tunneling barrier $A_{\rm max}$, the longer  the oscillation period.
On its way back, the wave packet broadens more and more due to reflections at the central barrier. 
Finally, after one period $t \simeq 3.6$, Fig.~\ref{Figure11}  (d), a large fraction is again located in the vicinity of the saddle-point, which, 
subsequently, splits once more.

In contrast, for a long ramping time $T_{\rm ramp}=30$, see Fig.~\ref{Figure11} (e--h), 
the wave function has enough time to adapt to the new boundary conditions,  
such that it rather smoothly follows the minima of the dynamically created double-well potential. The dynamics are still garnished, for this long but finite ramping time, by excitations of the first band, as identifiable by additional nodal structures in Fig.~\ref{Figure11} (g,h).

\begin{figure}[t]
	\includegraphics[width=\figwidthsingle\columnwidth]{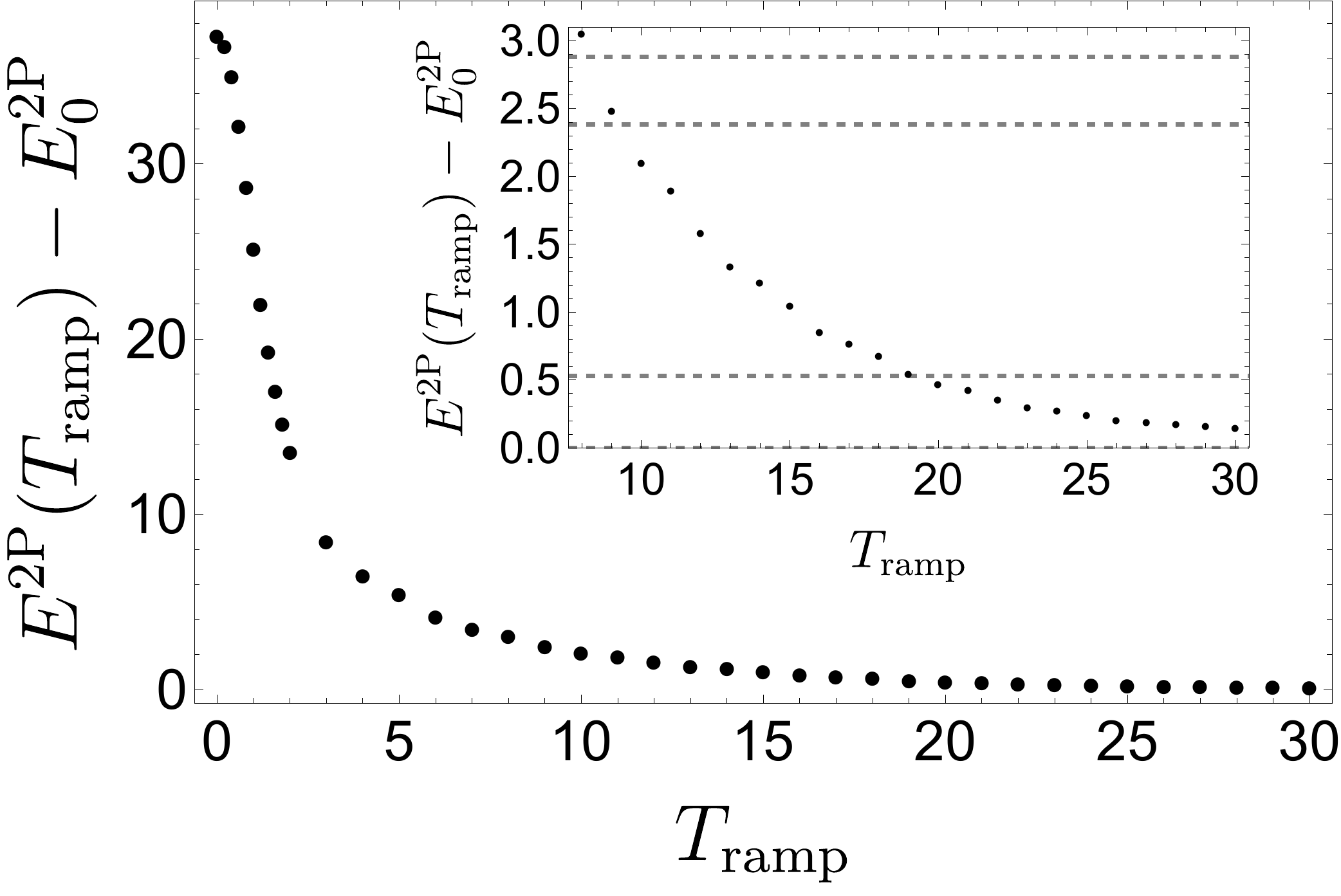}
	\caption{Two-body  energy expectation, Eq.~\eqref{2P_energy_expectation}, versus ramping time, after a fixed evolution time $t_0=200$, for $A_{\rm max}=30$ and $\lambda=1$. 
	The inset zooms into the range $T_{\rm ramp} \ge 8$, where the horizontal dashed lines indicate the low-lying eigenenergies of Eq.~\eqref{MB_Hamiltonian}, computed by FGH.
	FGH parameters: $x_{\rm max}=-x_{\rm min}=40$, $N_{\rm cut}=330$, and $N_{\rm Grid}=2047$;
	\textsc{Mctdh-x} parameters employed for the time-propagation:  $x_{\rm max}=-x_{\rm min}=12$, $N_{x}=512$, and $M=16$.
	\label{Figure12}   }
\end{figure}

Comparison of the nodal structures of the two-particle densities observed in Fig.~\ref{Figure11} for $T_{\rm ramp}=0$ and for $T_{\rm ramp}=30$, respectively, suggests that less energy is absorbed by the center of mass degree of freedom in the latter case (as expressed by considerably fewer nodal lines, indicative of smaller momenta). To corroborate this conjecture (which is based on evidence of short time dynamics only), we plot the two-particle energy expectation value 
\begin{align}
E^{2P}(T_{\rm ramp})=\langle \Psi(t_0)| \mathcal{ H^{\rm 2P}}(T_{\rm ramp})|\Psi(t_0) \rangle\, ,
\label{2P_energy_expectation}
\end{align} 
at $t_0=200\gg T_{\rm ramp} $, for variable $T_{\rm ramp} \in [0,30]$, in Fig.~\ref{Figure12}.
We observe a quick initial drop of the energy followed by a long tail  approaching smoothly the energy of the ground state, 
for $A_{\rm max}=30$ and  $\lambda=1$, i.e., $E^{2P}_0\simeq11.34$. 
The inset zooms  into the range  $T_{\rm ramp} \in [7.5,30.5]$, 
where the horizontal dashed lines indicate the eigenenergies of the two-particle system, with $A_{\rm max}=30$ and $\lambda=1$. 
The evolution of $E^{2P}(t_0)$ implies that, for $T_{\rm ramp} \geq19$, 
only transitions between the ground state and the first degenerate (recall Fig.~\ref{Figure6}) excited states occur. 
Thus, indeed, (quasi-)adiabatic switching does perform essentially no work on the many-particle system.

\begin{figure}[t]
	\includegraphics[width=\figwidthsingle\columnwidth]{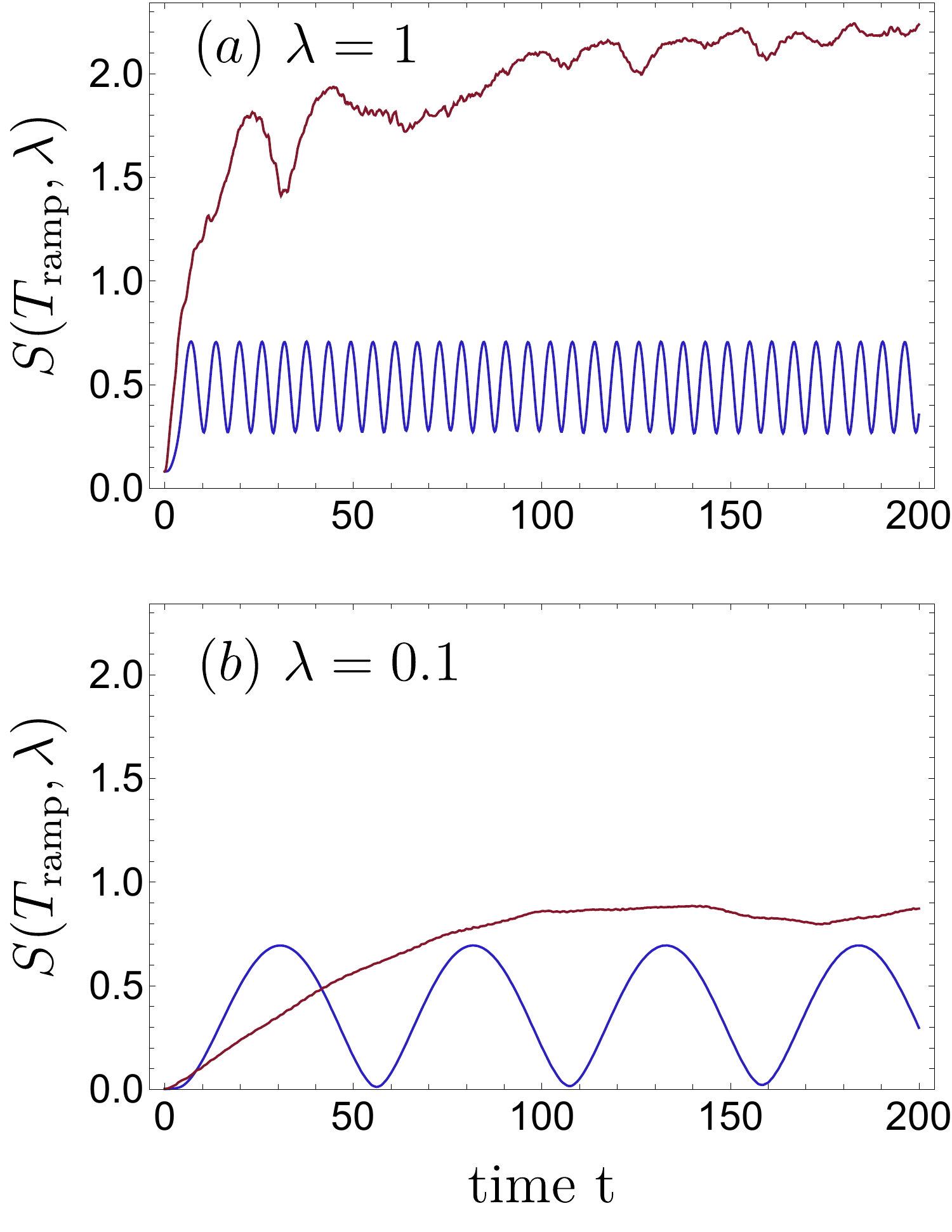}
	\caption{(Color online) 	
	Von Neumann entropy \eqref{Von_Neumann_Entropy} of the interacting two-particle state launched in the harmonic oscillator (interacting) two-particle ground state, 
	as a function of time, for 
	short and long ramping times, $T_{\rm ramp}=0.001$ (red)  and $T_{\rm ramp}=30$ (blue), and strong ($\lambda=1$, (a)) and 
	weak ($\lambda=0.1$, (b)) interaction, respectively. 
	For small $T_{\rm ramp}$, the entropy increases and finally saturates,
	whereas it oscillates for long ramping times.
    \textsc{Mctdh-x} parameters:  $x_{\rm max}=-x_{\rm min}=12$, $N_{x}=512$, and $M=16$.
	}
	\label{Figure13}   
\end{figure}

The static double well's entropy of the reduced single-particle density matrix increases from zero at $\lambda=0$ and saturates at $\ln2$ \cite{FS_thesis_2018, Murphy_Busch_2007, Murphy_McCann_2008}
with our definition \eqref{Von_Neumann_Entropy}
for $\lambda \to \infty$, $\forall A_{\rm max}$.
In our present, dynamical scenario --  where the harmonic trap is split into a double-well during a  time $T_{\rm ramp}$ -- we also expect 
the entropy to increase with the interaction. 
Figure~\ref{Figure13} shows the time-evolution of the entropy for two ramping times (red/blue) and for two values of the interaction strength, (a) $\lambda=1$ and (b) $\lambda=0.1$.

For short ramping time,  $T_{\rm ramp}=0.001$ (red lines), the entropy  increases  and saturates at $\approx 2.51$ which is well below the maximal value $S^{\rm max}=\log(M)\approx 2.77$ and which we verified with respect to the time evolution for $T_{\rm ramp}=0$ using the spectral decomposition based on our FGH computations from Section~\ref{sec_3}.
In agreement with the asymptotic behaviour of the energy expectation value observed in Fig.~\ref{Figure12}, the entropy oscillates with a single frequency for large ramping time, 
e.g.,  $T_{\rm ramp}=30$ (blue lines in Fig.~\ref{Figure13}). 
The stronger the interaction, the larger the frequency as well as the offset of the minima of the entropy oscillations.

\begin{figure*}[t]
	\includegraphics[width=\figwidthdouble\columnwidth]{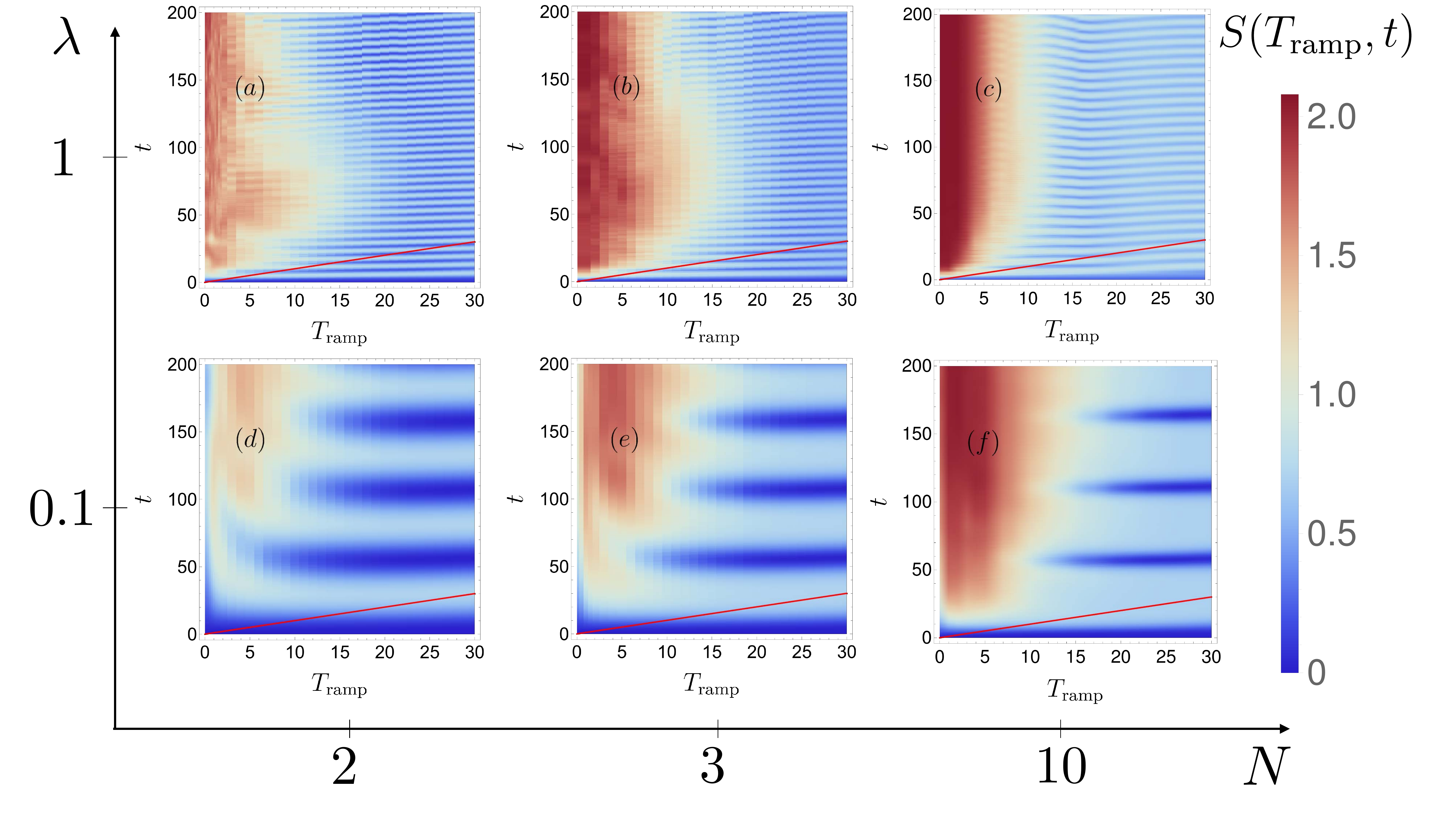}
	\caption{(Color online) Time evolution of the von Neumann entropy $S(T_{\rm ramp}, t)$, Eq.~(\ref{Von_Neumann_Entropy}),
	as a function of the ramping time $T_{\rm ramp}$, for a final barrier height $A_{\rm max}=30$, increasing particle number $N=2, 3, 10$ (from left to right),
	and interaction strengths
	$\lambda=1$ (a--c;) and $\lambda=0.1$ (d--f).
	The red line indicates the full switching duration $t=T_{\rm ramp}$ for the ramp to reach its maximum.
	(Parameter values employed in the \textsc{Mctdh-x} calculation: $x_{\rm max}=-x_{\rm min}=12$, $N_{x}=512$,  $M=8$.)
	}
	\label{Figure14}   
\end{figure*}

Monitoring the time evolution of the entropy over a broad interval of $T_{\rm ramp}$ allows us to map out the different dynamical regimes for  two bosons with $\lambda=1$ and $\lambda=0.1$, respectively, see Figs.~\ref{Figure14} (a) and (d). 
In full agreement with what we observed above for the dependence of 
the  energy expectation value on $T_{\rm ramp}$, the transition 
from diabatic to (quasi-)adiabatic dynamics is also here the primary feature:
For short ramping times, the entropy rapidly saturates at its equilibrium value, whereas, for a sufficiently slow ramp, an oscillation emerges, with a single, well-defined frequency (and decreasing amplitude, for increasing $T_{\rm ramp}$).
A discrete Fourier transform of the signal for 
$ T_{\rm ramp}\approx 19-30$
shows  that this frequency is determined by the energy gap (see Section \ref{sec3_subA}) between the ground and first excited state,
\begin{align}
\nu(\lambda)=\frac{E_1^{\rm 2P}(\lambda)-E_0^{\rm 2P}}{\pi}\, .
\end{align}
Indeed,
only two eigenstates of the reduced single-particle density matrix -- with opposite parity and densities which closely resemble the typical structure of the double-well ground state doublet \cite{FS_thesis_2018} --
contribute to the dynamics in this oscillating region (not shown). 
For intermediate ramping times
($ T_{\rm ramp}\approx 10-19$), the structures observed in Fig.~\ref{Figure14} (a, d) still express the switching-induced, coherent coupling of more than just two interacting eigenstates, because in this regime the dynamics are not yet (quasi-) adiabatic (in agreement with our discussion of Fig.~\ref{Figure12}).\\

Remarkably, although the detailed spectral structures are rather different for two and three particles 
(see Figs.~\ref{Figure3} and \ref{Figure4}), the ramping-induced time-dependence 
of the von Neumann entropy is qualitatively similar for $N=2, 3$, and even $N=10$ (where we cannot access the spectral structure, with our presently available
numerical resources)
see Fig.~\ref{Figure14} (a,b,c), for $\lambda=1$, and 
Fig.~\ref{Figure14} (d,e,f), for $\lambda=0.1$. We attribute this feature to the coarse-graining effect of a diabatic switch, where only the 
effective density of states has to be gauged against the spectral width of the time-dependent perturbation.
Closer inspection suggests that efficient excitation is achieved for slightly longer switching times with increasing particle number,
which is consistent with the increase of the density of states with $N$.
The frequency $\nu \sim E_1^{\rm NP} -E_0^{\rm NP}$ of the  entropy oscillations slowly decreases with the number of particles,
since the energy gap $\Delta E = E_1^{\rm NP}-E_0^{\rm NP}$ between the first excited state  and the 
ground state  decreases with $N$, i.e., $\Delta E (N=2)>\Delta E(N=10)$. 
Also note that the oscillating regime of Fig.~\ref{Figure14} (c, f) corresponds to the twofold fragmented BEC discussed in Refs.~\cite{cederbaum_splitting_dynamics, Menotti_2001, Sakmann_Cederbaum_PRL_2009}.
Similar results are observed for different barrier heights (not shown) \cite{FS_thesis_2018}.

Let us conclude this section with a remark on the convergence of the \textsc{Mctdh-x} results reported in Fig.~\ref{Figure14} (c,f): 
For moderate and large $T_{\rm ramp}\gtrsim7$, only two orbitals of the employed $M=8$ orbitals have a significant population and the entropy $S$ remains significantly smaller than the maximal value $S^{\rm max}=\log(M)$. 
From this fact it can be inferred that the wave function is accurately described in these \textsc{Mctdh-x} computations at $T_{\rm ramp}\gtrsim7$.
However, for small ramping times ($T_{\rm ramp}\lesssim 7$) all $M=8$ employed orbitals in the computation were populated. 
Consequently, the entropy reaches its maximum $S^{\rm max}$. 
This maximal entropy for small $T_{\rm ramp}$ implies that the Hilbert space provided by \textsc{Mctdh-x} is not large enough to host the complete dynamics of the many-body wave functions and more orbitals ($M>8$) would therefore be necessary to achieve convergence.
While the quantitative behaviour of the entropy $S(T_{\rm ramp},t)$ at small $T_{\rm ramp}\lesssim 7$ in Fig.~\ref{Figure14} (c,f) therefore cannot be considered 
fully converged, the observed behaviour is qualitatively equivalent to that resulting for smaller particle numbers, where convergence
of \textsc{Mctdh-x} could be achieved with a smaller number $M=2,4,6$ of orbitals, and is also consistent with our FGH-based analysis for $N=2$
particles (see Fig.~\ref{Figure12}). This suggests that the results reported in Fig.~\ref{Figure14} (c,f) correctly indicate the  qualitative trend of the evolution also for short ramping times.

\section{Conclusions}
\label{sec_5}

We analyzed the spectral structures and the dynamics of few interacting bosons in a one-dimensional double-well potential, 
for both a static and a time-dependent potential barriers, beyond the two-mode approximation.
To this end, we used  three complementary numerical methods. 
The Fourier Grid Hamiltonian method was employed to extract the full spectral information for two interacting bosons, whereas a Bose-Hubbard representation of the continuous double-well potential was found to be more efficient to describe the spectral structure of the  three-particle case.
Furthermore, we used the \textsc{Mctdh-x} method to simulate the dynamical evolution of $N=2, 3$, and 10 interacting bosons in a potential 
with time-dependent barrier strength.

Our spectral analysis highlights the dependence of the energy spectrum on the interaction strength, on the one hand, and on the potential barrier height,
on the other.
Ramping up a barrier in the center of an initially harmonic potential introduces a metamorphosis of state space from 
a simple, highly degenerate harmonic oscillator progression, into a sequence of states which exhibit the characteristic degeneracies associated with 
tunneling between symmetric wells, below the barrier energy, and a harmonic-like spectrum sufficiently high above the barrier, separated by a 
range around the barrier energy which mediates between both classes. 
Interactions lift many of the energetic degeneracies and eventually induce mixing of energetic manifolds which otherwise remain well-separated.

While for two (on-site interacting) particles distributed over two (deep) wells eigenstates exist which remain unaltered by finite interactions,
this is no more true for three particles in the same potential, since at least two particles then have to interact: 
two manifolds of states emerge corresponding to states where two or three particles are interacting. 
We supplemented our spectral analysis by inspecting many-particle probability densities in configuration space, 
which directly exhibit the spatial correlations inscribed into the many-body tunnelling dynamics, for both energetically low- and high-lying states. 
For three particles, we visualised the loss of information about correlations when tracing from the three-body density to the two-body, and, eventually, to the one-body density. 

We used that spectral information to decipher the tunnelling dynamics of two interacting particles in a static double-well. 
In particular, we compared and characterized Josephson oscillations of two interacting bosons 
prepared in a superposition of excited states with energies either well below or close to the potential's saddle point.
Inspection of the expansion coefficients of the evolved two-particle state in the interacting two-particle basis provided evidence 
that a simple three-level description of the dynamics fails at sufficiently strong interactions.
The Josephson period at energies close to the saddle-point
is much smaller and robust with respect to the interaction.
In agreement with observations in Ref.~\cite{Folling_Bloch_Nature_2007}, we confirm a second-order, pairwise tunnelling process.

Finally, we investigated the spreading behaviour of the many-particle state, when initially prepared in the many-particle harmonic oscillator
ground state, under diabatic vs. (quasi-) adiabatic switching of a central barrier -- transforming the potential into a double well. Diabatic switching 
leads to efficient energy transfer through the population of a large number of many-particle excited states, as quantified by the time-evolution of the 
von Neumann entropy of the reduced single-particle density matrix, while a (quasi-) adiabatic ramp only populates weakly excited states. This phenomenology 
emerges already for two interacting particles and -- due to the increasing spectral density -- gets more pronounced for ten particles, the largest particle 
number here considered. 

\FloatBarrier  
\begin{acknowledgments}
We would like to thank Gabriel Dufour for inspiring discussions and a careful reading of the manuscript. 
We acknowledge support through the EU collaborative Project QuProCS (Grant Agreement 641277), as well as by the state of Baden-W\"urttemberg through bwHPC (NEMO and JUSTUS clusters).
FS acknowledges financial support by the Swiss National Science Foundation (SNSF) and the NCCR Quantum Science and Technology.
LdFdP  acknowledges the Alexander von Humboldt-Foundation for financial support.
MABM acknowledges financial support from CONACyT postdoctoral fellowship program. AUJL acknowledges financial support by the Austrian Science Foundation (FWF) under grant P32033 and computation time at the Hazel Hen cluster at the HLRS Stuttgart.
\end{acknowledgments}

\bibliography{dw_bibliography}

\appendix
\section{Fourier Grid Hamiltonian Method \label{Method_FGH}}

The FGH method \cite{Marston_FGH_1989, Balint_Kurti_1991}  is a special case of a discrete variable representation where the eigenfunctions of the single-particle Hamiltonian are computed directly as the amplitudes of the wave function on the grid points. 
The results of the FGH method -- the single-particle eigen-energies and -states -- are then used as a basis set representation of the many-particle Hamiltonian. A subsequent exact diagonalization determines the many-body spectrum.

The FGH numerical implementation requires a discretization of the continuous coordinate space 
by a discrete set of an odd number of $N_{\rm Grid}$ lattice points distributed in a uniform manner,
such that $x_m=x_{\rm min}+ m \Delta x$, with $m \in[0, N_{\rm Grid}-1]$.
This discretization leads to a grid and momentum spacing
\begin{eqnarray}
\Delta x &=& \frac{x_{\rm max}-x_{\rm min}}{N_{\rm Grid} }~, \\
\Delta p &=& \frac{2\pi}{x_{\rm max}-x_{\rm min}}~. 
\label{FGH_grid_space_momentum}
\end{eqnarray}
From Eq.~\eqref{MB_Hamiltonian}, the single-particle Hamiltonian matrix elements  $\mathcal { H}_{mn} = \langle x_m |   \mathcal {   H}  | x_n \rangle  $ read  \cite{Marston_FGH_1989}
\begin{eqnarray}
\nonumber
  \mathcal { H}_{mn} &=&\sum_{l=1}^{\frac{N_{\rm Grid} -1}{2}} \frac{(l \Delta p)^2}{\Delta x N_{\rm Grid}}  \cos{\left( \frac{2\pi l(m-n)}{N_{Grid}} \right )} \\
  &+& \frac{V(x_m,t)}{\Delta x} \delta_{mn} , \ \ \ \ \ 
\label{FGH_grid_spacing}
\end{eqnarray}
with the potential $V(x_m,t)$ defined by Eq.~\eqref{Double_Well_Potential}.
Using this discretized procedure,  the wave function may  be  represented 
as  a vector  on  a  discretized  grid  of  points
\begin{eqnarray}
| \Psi \rangle =  \Delta x   \sum_m \psi_m |x_m\rangle, 
\label{FGH_wave_function}
\end{eqnarray}
with $\psi_m = \psi(x_m)=\langle x_m | \Psi \rangle $ the value of the wave function evaluated at  $x_m$, and 
with orthogonality  condition $\Delta x  \langle x_m | x_n \rangle = \delta_{mn}$.

We thus obtain a discretized position representation of the single-particle Hamiltonian and must compute the 
eigenvalues of this Hamiltonian matrix. To this end, we consider the energy expectation value with respect to  state $| \psi \rangle$, given by
\begin{equation}
E=\frac{\langle \Psi |   \mathcal {  H}   | \Psi \rangle}{\langle \Psi  | \Psi \rangle}
=  \frac{\sum_{mn}\psi_m^*    \Delta x \mathcal { H}_{mn}  \psi_n}{  \sum_{m}|\psi_m |^2}.
\label{FGH_energy}
\end{equation}
The minimization of this energy functional  
 by variation of the coefficients $\psi_m$ leads to the secular equations
\begin{eqnarray}
\sum_{n=0}^{N_{\rm Grid}-1} \left[ \Delta x   \mathcal { H}_{mn} - E^{1P}_m \delta_{mn }\right]   \psi_m = 0, \label{FGH_single_particle_eigenbasis} \\
\nonumber
 m=0,\dots,N_{\rm Grid}-1, \qquad \qquad \qquad \qquad
\end{eqnarray}
with eigenvalues $E^{1P}_m$.
The  eigenvectors $| \Psi_m \rangle $  give  directly  the (approximate)  values  of  the  solutions  of  the  Schr\"odinger 
equation  evaluated  at  the  grid  points.  As discussed below, the convergence of the method in the absence of free scattering states 
is, a posteriori, well controlled, thus leading to a \textit{numerically exact} result, i.e., 
with an error of the order of the machine precision. 
Furthermore, since the single-particle Hamiltonian is real and symmetric, these eigenstates can always be chosen to be real.
Note that the double-well potential investigated does not exhibit free scattering solutions, but only bound states. 

The precision of the FGH method can be enhanced by varying two characteristic  parameters: 
(1) The range $ x_{\rm max}-x_{\rm min}$ determines the maximum value of the potential $V_{\rm max}(x_{\rm max}, t)$.
	As soon as the energy of a given bounded state $| \Psi_n \rangle $  does not exceed the truncation $V_{\rm max}(x, t)$, convergence can be controlled.
	For instance, with $x_{\rm max}=-x_{\rm min}=40$ and $A_{\rm max} \lessapprox 30$, roughly  the $N_{\rm cut}=600$ lowest-lying energy eigenstates can be converged with a precision up to $10^{-9}$.
(2) Increasing the number of grid points $N_{\rm Grid}$  within a fixed range $x_{\rm max}-x_{\rm min}$ improves the 
	accuracy of the eigenenergies of the $N_{\rm cut}$ states toward the exact solutions. 
	Typically, with $x_{\rm max}=-x_{\rm min}=40$ and $N_{\rm cut}=330$, we used  $N_{\rm Grid}=2047$.
	These parameters ensure an energy convergence up to $10^{-9}$ in natural units and satisfy the orthonormality $\langle \Psi_n |\Psi_m \rangle=\delta_{n m}$  of the generated eigenstates, to double machine precision.

Using second quantization, the two-body Hamiltonian, expressed in terms of the single-particle eigenbasis obtained from the FGH method, reads 
\begin{equation}
 \mathcal {H}^{2P}   = \sum_{k=0}^{N_{\rm cut}-1} E_k^{1P} \hat{n}_k  +\frac{1}{2}\sum_{ksql=0}^{N_{\rm cut}-1} 
 W^{\phantom{\dagger}}_{ksql} \hat{a}^\dagger_k \hat{a}^\dagger_s \hat{a}_q^{\phantom{\dagger}} \hat{a}_l^{\phantom{\dagger}}~,
\label{Hamiltonian_FGH_2P}
\end{equation}
with $\hat{a}^\dagger_k $ ($\hat{a}_k^{\phantom{\dagger}}$) the creation (annihilation) operator in state $k$, and
where $\hat{n}_k = \hat{a}^\dagger_k \hat{a}_k^{\phantom{\dagger}} $ counts the number of particle in state $k$.
The  matrix element $W_{ksql}$ originates from contact interactions.
The Hamiltonian matrix,  of  dimension 
\begin{equation}
\text{dim}( \mathcal {  H}^{2P} )= \frac{N_{\rm cut} (N_{\rm cut}+1)}{2}~,
\label{Dimension_2P_hilbert_space}
\end{equation}
is computed in the  Hilbert space of  symmetrized and normalized
two-body states $| \psi_n \psi_m \rangle $, constructed from the single-particle product states such that
\begin{equation}
| \Psi_n \Psi_m \rangle \equiv \frac{ | \Psi_n \rangle \otimes  | \Psi_m  \rangle   +   | \Psi_m \rangle \otimes  | \Psi_n  \rangle  }{\sqrt{2} \sqrt{1+ \langle \Psi_n | \Psi_m \rangle }} ~,
\label{2P_exact_diagonal_basis}
\end{equation}
for $N_{\rm cut}\geq n\geq m \geq 1$.
Using this two-particle basis, the diagonal Hamiltonian matrix elements  read
\begin{equation}
\bra{\Psi_n \Psi_m}\sum_{k=0}^{N_{\rm cut}-1} E_k^{1P} \hat{n}_k\ket{\Psi_{n'} \Psi_{m'} } = ( E_n^{1P}+ E_m^{1P}) \delta_{nn'} \delta_{mm'} ,
\label{FGH_one_body_diagonal_terms}
\end{equation}
whereas the off-diagonal interacting terms read
\begin{align}
\nonumber
\frac{1}{2}\sum_{ksql=0}^{N_{\rm cut}-1} W_{ksql} \bra{\Psi_n \Psi_m} \hat{a}^\dagger_k \hat{a}^\dagger_s \hat{a}_q^{\phantom{\dagger}} \hat{a}_l^{\phantom{\dagger}}
\ket{\Psi_{n'} \Psi_{m'} }\\
=\begin{cases}
     W_{nnn'n'},  \ \ \  \ \  \text{for }n= m, n' = m', \\
    \sqrt{2} W_{nmn'n'}, \   \text{for }n \neq m, n' = m',\\
  \sqrt{2} W_{nnn'm'}, \  \text{for }n= m, n' \neq m', \\
  2 W_{nmn'm'},  \ \ \    \text{for }n \neq m, n' \neq m',
  \end{cases}
  \label{FGH_two_body_offdiagonal_terms}
\end{align}  
with 
\begin{eqnarray}
 \label{Wksql}
 W_{ksql}=\lambda \sum_{m=0}^{N_{\rm Grid}-1} \Delta x ~ \psi^k_m \psi^s_m \psi^q_m \psi^l_m 
\end{eqnarray}
numerically calculated using a Kahan summation algorithm \cite{Kahan} to minimize the accumulated numerical error.  

Then, the Hamiltonian matrix is diagonalized with \textsc{Mathematica}'s build-in \textsc{Lapack}-routines and MKL parallelization feature, 
which ultimately determine a number of $ \text{dim}( \mathcal {   H}^{2P}) $ eigenvalues $E_n^{2P}$ and associated eigenvectors $\ket{\Psi_n^{2P}}$.
The time evolution of the interacting two-particle system is given by the spectral decomposition 
\begin{equation}
\ket{\Psi(t)}=\sum_{n=0}^{  \text{dim}( \mathcal {  H}^{2P} ) -1  }  e^{-itE^{2P}_n}  \bra{\Psi_n^{ 2P}}  \Psi(t=0) \rangle ~\ket{\Psi_n^{ 2P}}, 
\label{FGH_Two_Particles_Time_Evolution_Wave_Function}
\end{equation}
with initial state $\ket{\Psi(t=0)}$.

\section{Bose-Hubbard model in the continuum \label{Method_BH}}

The discretization of the continuous configuration space as performed hereafter ultimately leads to a Hamiltonian 
which exhibits the familiar structure of a Bose-Hubbard Hamiltonian, 
amended by a site-dependent potential form. 
Therefore, the model developed below is referred to as the Bose-Hubbard (BH) model in the continuum \cite{Fleischhauer_2010_BHmodel_continuum}.
This approach gives access to the energy spectrum of two and three interacting bosons
with a good accuracy. The main advantage of this technique is that 
its convergence weakly depends on the interaction strength, which is not the case with the FGH method for which 
the matrix to diagonalize is dense in presence of interactions, then introducing high CPU time and memory costs.\\

Starting from the generic many-body Hamiltonian  for  $N$ ultracold particles  in the continuum limit, 
with contact interactions and double-well potential $V(x,t)$ (Eq.~\ref{Double_Well_Potential}),
\begin{eqnarray}
\nonumber
\mathcal{ H}^{NP} &=&\int \text{d}x ~\hat{\Psi}^\dagger(x)\left[-\frac{1}{2}\frac{\partial^2}{\partial x^2} +V(x,t)\right]\hat{\Psi}(x) \\
&+&\frac{\lambda}{2}\int \text{d}x  ~\hat{\Psi}^\dagger(x) \hat{\Psi}^\dagger(x) 
\hat{\Psi}(x)\hat{\Psi}(x),
\label{eq:many-body-hamiltonian-field-operators-hubbard}
\end{eqnarray}
we use the single-band description.
For practical implementation aspects, 
the continuous space is artificially discretized by covering it with Wannier functions. 
We  can then expand the field operators $ \hat{\Psi} ( x) $ in the basis of localized and orthonormal Wannier functions of the lowest-lying band $w_0 ( x-x_i)$:
\begin{eqnarray}
\hat{\Psi}({ x}) = \sum_{i=1}^{L} \hat{a}_i w_0 ( x-x_i), 
\label{BH_Wannier}
\end{eqnarray}  
with  $\hat{a}_i$ the annihilation operator for  a particle  in the single-mode Wannier function $w_0 ({ x-x_i})$ at site $i$, 
and  $L$  the number of sites in the discretization (assimilable to the number of grid points in the FGH method). 
Inserting the expansion \eqref{BH_Wannier} in Eq.~\eqref{eq:many-body-hamiltonian-field-operators-hubbard}, we obtain
 \begin{eqnarray}
\mathcal{ H}^{NP}  & = &-\frac{1}{2}\sum_{i j}  \int \text{d} x\   \hat{a}^\dagger_i w^\star_0 ({ x-x_i})   \frac{\partial^2}{\partial x^2}  \hat{a}_j w_0 ({ x-x_j}) \notag\\
&+& \sum_{i} \hat{n}_i    V(x_i, t) \notag\\
&+& \frac{\lambda}{2}  \sum_{i}\hat{n}_i(\hat{n}_i-1) \int \text{d} x   \ | w_0 ({ x-x_i}) |^4 ~,
\label{eq:models:zwischen}
\end{eqnarray} 
where  $\hat{n}_i= \hat{a}^\dagger_i \hat{a}_i^{\phantom{\dagger}}$  and  
$\hat{n}_i(\hat{n}_i-1)= \hat{a}^\dagger_i \hat{a}^\dagger_i \hat{a}_i^{\phantom{\dagger}} \hat{a}_i^{\phantom{\dagger}}$.

Then, the kinetic term is discretized by a finite lattice spacing $\delta_x=(2 x_{\rm max})/({L-1})$, 
\begin{eqnarray}
\frac{\partial^2}{\partial x^2}  \hat{a}_j w_0 ({ x-x_j}) &\simeq& \frac{1}{\delta_x^2}     \hat{a}_{j+1} w_0 ({ x-x_{j+1}}) \\
&+&\frac{1}{\delta_x^2} \hat{a}_{j-1} w_0 ({ x-x_{j-1}})  \\
&-& \frac{2}{\delta_x^2} \hat{a}_{j} w_0 ({ x-x_{j}}) ~,
\label{eq:tight-binding-approximation-BH}
\end{eqnarray}  
such that $L$ grid points are uniformly distributed between $x_{\rm min}=-x_{\rm max} $ and $x_{\rm max} $, 
and the discretized Wannier function reads
\begin{equation}
w_0 ({ x-x_i})  \to  w_{0i} / \sqrt{\delta_x}.
\end{equation}  
With the on-site interaction strength
\begin{eqnarray} 
U \equiv \lambda  \sum_i  \ | w_{0i} |^4,
\label{eq:Uvslambda-BH}
\end{eqnarray}  
the BH Hamiltonian in the continuum takes the final form
\begin{eqnarray}
\mathcal{ H}^{NP}   &=& -\frac{1}{2 \delta_x^2}    \sum_{i=1}^{L-1}  \left(  \hat{a}^\dagger_i  \hat{a}_{i+1}^{\phantom{\dagger}}  + \hat{a}^\dagger_{i+1}  \hat{a}_{i}^{\phantom{\dagger}}  \right )  \notag\\
&+& \sum_{i=1}^{L}   \left( V_i(t)  +\frac{1}{ \delta_x^2} \right) \hat{n}_i  \notag\\
&+&\frac{U}{2 \delta_x}  \sum_{i=1}^{L}  \hat{n}_i(\hat{n}_i-1) ~.
\label{BH_Hamiltonian_final}
\end{eqnarray} 
The double-well potential, in accordance with Eq.~\eqref{Double_Well_Potential},  is then encoded by the explicit form 
\begin{eqnarray}
V_i(t)= \frac{x_i^2}{2}+A(t)e^{-x_i^2/2}, \notag\\
x_i \in \{-x_{\rm min}, -x_{\rm min}+\delta_x, \dots, x_{\rm max}\}
\label{BH_discretized_double_well}
\end{eqnarray}
which, for $A(t)=0$, turns into harmonic (single-well) trapping potential.
Using the Fock basis $\ket{\vec{n}}$, the Hamiltonian matrix elements read
\begin{eqnarray}
\mathcal{H}^{NP}_{mn}  = \bra{\vec{m}} \mathcal{ H}^{NP} \ket{\vec{n}}.
\label{BH_matrix_elements}
\end{eqnarray}
For three particles, the matrix to diagonalize has a size of 
\begin{eqnarray}
\text{dim}( \mathcal {   H^{\it 3P}} )= \frac{1}{6}L(L+1)(L+2).
\label{Dimension_3P_hilbert_space}
\end{eqnarray}

Despite the sparsity of the matrix -- which is a great advantage compared to the FGH method  -- 
the diagonalization of this matrix is rather challenging.
Indeed, for 3 particles, we have used $x_{\rm max}=-x_{\rm min}=10$ and $L=231$, leading to a matrix size of  $2,081,156\times2,081,156$.
To obtain parts of the spectrum with reliable degeneracies, 
we used the  \textsc{Mathematica}'s implementation of the \textsc{Feast} eigensystem solver \cite{feast_description} for sparse matrices, which is inspired by the contour integration and density matrix representation in quantum mechanics \cite{QM_polizzi_feast}. 
Within a given energy search interval $\{E_{\rm min}, E_{\rm max}\}$, the \textsc{Feast} algorithm reduces the size of the eigenvalue problem 
to a subspace of size associated to the number of eigenvalues in this interval.
This approach  naturally captures the degeneracies in the energy spectrum \cite{feast_description}. 
Moreover, using  \textsc{Mathematica}, the \textsc{Feast} method is  MPI parallelized over all processors on a single node on the cluster. \\

\section{Multiconfigurational Time-Dependent Hartree method for indistinguishable particles \label{Method_MCTDH}}

\textsc{Mctdh-x} allows for the investigation of interacting particles in many scenarios, e.g., 
interacting bosons or fermions in optical lattices \cite{axel_hubbard}, quantum vortex re-connections in a Bose-Einstein condensate  \cite{Wells2015}, 
or  bosons in double-well potentials \cite{zollner_few_bosons,  zollner_Schmelcher_PRA_2006,  cederbaum_splitting_dynamics}. 
In our context, this method is useful for the investigation of $N$ interacting bosons in a time-dependent double-well potential.
Nevertheless, this method is not efficient for the calculation of the entire energy spectrum, thus justifying our use of the FGH and BH methods for few particles.
In the following, we outline the basic steps towards the \textsc{Mctdh-x} equations of motion, see
Ref.~\cite{cederbaum_MCTDHX} for supplemental details regarding the method.

The aim is to solve the time-dependent Schr\"odinger equation
\begin{eqnarray}
i \frac{\partial}{\partial t } \ket{\Psi} =   \mathcal {  H}  \ket{\Psi}~,
\label{MCTDH_time_dependent_Schodinger_equation}
\end{eqnarray}
with many-body Hamiltonian $  \mathcal {  H} $ defined by Eq.~\eqref{MB_Hamiltonian}.
To do so, we first formulate a general multiconfigurational ansatz for the wave function based on 
truncating the field operator
\begin{eqnarray}
\hat{\Psi}(x, t)= \sum_{k} \hat{a}_{k}(t) \phi_k(x, t) 
\label{MCTDH_field_operators1}
\end{eqnarray}
from an infinite to a finite sum of $M$ operators, i.e.,
\begin{eqnarray}
\hat{\Psi}(x, t)\approx \sum_{k=1}^{M} \hat{a}_{k}(t) \phi_k(x, t) ~.
\label{MCTDH_field_operators}
\end{eqnarray}
Under this assumption,  the bosonic ansatz for the many-body wave function reads
\begin{eqnarray}
\ket{\Psi } = \sum_{\{\vec{n} \}} C_{\vec{n}} (t)  \prod_{k=1}^{M}   \frac{(\hat{a}^\dagger_{k}(t))^{n_k}}{\sqrt{n_k!}} \ket{\rm vac} ,
\label{eq:MCTDHX-state}
\end{eqnarray}
where the summation runs over all (symmetrized) basis states of the Hilbert space.
The vector  $\vec{n} = (n_1, n_2, \dots, n_M)$ 
represents the occupations of the orbitals that preserve the total number of particles
$n_1+n_2+n_3+ \dots+ n_M=N$,  $M$ is the number of orbitals $\phi_k(x, t) $, and $ \ket{\rm vac}$
is the vacuum.
This (a posteriori controlled) assumption, which is the key idea of \textsc{Mctdh-x}, 
greatly reduces the computational effort.

Using this ansatz, the time-dependent Schr\"odinger equation is solved by using the 
time-dependent variational principle for minimizing the action functional \cite{MCTDH_Variational_Principle}
\begin{widetext}
\[
\mathcal {  S} \big[ \{  C_{\vec{n}}(t)\} ,\{\phi_k(x,t)\}   \big]   =
\int \text{d}t~\bigg[  \bra{\Psi (t) }  \left (  \mathcal {  H} - i \frac{\partial}{\partial t } \right)  \ket{\Psi (t) } -\sum_{k,j=1}^{M}\mu_{kj}(t)  \bigg (   \langle \phi_k(t)   \ket{ \phi_j(t)}-\delta_{kj}\bigg)    \bigg],
\]
\end{widetext}
where  the time-dependent Lagrange multipliers $\mu_{kj}(t) $  enforce the orthonormality of the orbitals.

The minimization of the action  $\mathcal {S}$ finally leads to the \textsc{Mctdh-x} equations of motion, i.e.,
a coupled set of first-order non-linear differential equations  \cite{cederbaum_MCTDHX}
\begin{eqnarray}
i\frac{\partial}{\partial t}C_{\vec{n}}(t) &=&\sum_{\vec{m}}\bra{\vec{n},t}    \mathcal { H}    \ket{\vec{m},t}  C_{\vec{m}}(t) \label{MCTDHX_coeff}~,\\
\nonumber
i\frac{\partial}{\partial t} \ket{\phi_k} &=& \mathbf{P}  \bigg [    \left (  - \frac{1}{2}   \frac{ d^2 }{d x^2} +V(x,t) \right )    \ket{\phi_k}\\
&+&\lambda \sum_{\alpha\beta\gamma \delta}^{M}
\{ \rho^{(1)} \}^{-1}_{k\alpha}    \rho^{(2)}_{\alpha\beta\gamma \delta}   \phi^*_\beta (x,t) \phi^{\phantom{*}}_\delta (x,t) \ket{\phi_\gamma}  \bigg ],  \ \ \ \ \  \ \
\label{MCTDHX_equation_motion}
\end{eqnarray}
where $\mathbf{P}= 1-\sum_{j=1}^{M} \ket{\phi_j}\bra{\phi_j} $ denotes the projection operator, 
and where $\rho^{(1)}_{k\alpha} = \bra{\Psi } \hat{a}^\dagger_{k}\hat{a}^{\phantom{\dagger}}_\alpha  \ket{\Psi }$ 
and $\rho^{(2)}_{\alpha\beta\gamma \delta} = \bra{\Psi } \hat{a}^\dagger_{\alpha}  \hat{a}^\dagger_{\beta}  \hat{a}^{\phantom{\dagger}}_\gamma \hat{a}^{\phantom{\dagger}}_\delta   \ket{\Psi } $ are respectively 
the matrix elements of the  reduced single-  and two-particle density matrices.
The projector $\mathbf{P}$ vanishes \textit{exactly} only in the limit $M \to \infty$, 
thus Eq.~\eqref{MCTDHX_coeff} becomes equivalent to the time-dependent Schr\"odinger equation.
On the other side, the \textsc{Mctdh-x}  method with one orbital,  i.e., $M=1$, is equivalent to the  Gross-Pitaevskii mean-field
where only  one coefficient  $C_{0,0,..,N,..,0}(t)$ contributes.
Therefore, the accuracy of \textsc{Mctdh-x} strongly depends on the choice of the number of orbitals $M$ 
used in the simulations and  the  convergence of the \textsc{Mctdh-x}  results  can be improved by increasing the number of orbitals $M$ \cite{error_mctdhx, MCTDHX_Lode_Streltsov_2012, axel_hubbard, fasshauer2016multiconfigurational}.

We have used the freely available software implementation \cite{softwareMCTDHX}
where the spatial discretization relies on a discrete variable representation (DVR) combined 
with a fast Fourier transformation \cite{beckMCTDHX}.
In practice, we have used $M \in \{8, 20\}$ orbitals, $x_{\rm max}=-x_{\rm min}=12$ and ${N}_{x}=512$ grid points. 
With these parameters employed in \textsc{Mctdh-x}, 
the absolute error of the eigenenergies -- computed by improved relaxation -- for two interacting particles in a harmonic trap, with respect to the exact ones, is found to be at the order of $10^{-4}$--$10^{-2}$.
See Refs.~\cite{error_mctdhx, axel_hubbard, MCTDHX_Lode_Streltsov_2012, fasshauer2016multiconfigurational} for more details about the convergence of the method.

\end{document}